\documentclass[%
 aip,
 amsmath,amssymb,
 reprint,%
]{revtex4-1}
\usepackage{amsmath,amssymb,amsthm}
\usepackage[capbesideposition=right]{floatrow}
\usepackage{graphicx}
\usepackage{dcolumn}
\usepackage{bm}

\usepackage[utf8]{inputenc}
\usepackage[T1]{fontenc}
\usepackage{mathptmx}
\usepackage{etoolbox}

\makeatletter
\def\@email#1#2{%
 \endgroup
 \patchcmd{\titleblock@produce}
  {\frontmatter@RRAPformat}
  {\frontmatter@RRAPformat{\produce@RRAP{*#1\href{mailto:#2}{#2}}}\frontmatter@RRAPformat}
  {}{}
}%
\makeatother

\usepackage{hyperref}
\usepackage{amsthm}
\usepackage{graphicx}
\usepackage{subfigure} 
\usepackage{color,bbm,caption}
\usepackage[normalem]{ulem}

\theoremstyle{plain}

\theoremstyle{definition}

\begin{document}

\title[Reliability and robustness of oscillations in some slow-fast chaotic systems]{Reliability and robustness of oscillations in some slow-fast chaotic systems}
\author{Jonathan Jaquette}
\affiliation{Department of Mathematics and Statistics, Boston University}
\affiliation{Department of Mathematics, Brandeis University.}
\affiliation{Volen National Center for Complex Systems, Brandeis University.}
\author{Sonal Kedia}
\affiliation{Volen National Center for Complex Systems, Brandeis University.}
\affiliation{Biology Department, Brandeis University.}
\author{Evelyn Sander}
\affiliation{Department of Mathematical Sciences, George Mason University}
\author{Jonathan D. Touboul}
\email{jtouboul@brandeis.edu}
\affiliation{Department of Mathematics, Brandeis University.}
\affiliation{Volen National Center for Complex Systems, Brandeis University.}

\date{\today}%

\begin{abstract}
	A variety of nonlinear models of biological systems generate complex 
	chaotic behaviors that contrast with biological homeostasis, the observation
	that many biological systems prove remarkably robust in the face of 
	changing external or internal conditions. Motivated by the subtle dynamics
	of cell activity in a crustacean central pattern generator (CPG), this paper proposes a
	refinement of the notion of chaos that reconciles homeostasis and chaos
	in systems with multiple timescales. We show that systems displaying relaxation 
	cycles while going through chaotic attractors generate chaotic dynamics that are regular at
	macroscopic timescales and are thus consistent with physiological function.
	We further show that this relative regularity may break down through global bifurcations 
	of chaotic attractors such as crises, beyond which the system may also generate erratic
	activity at slow timescales. We analyze these phenomena in detail in
	the chaotic Rulkov map, a classical neuron model known to exhibit 
	a variety of chaotic spike patterns. This leads us to propose that the passage of slow 
	relaxation cycles through a chaotic attractor crisis is a robust, general mechanism 
	for the transition between such dynamics. We validate this numerically in three other models:
	a simple model of the crustacean CPG neural network, a discrete cubic map, and a continuous flow.  
	
\end{abstract}

\maketitle

\begin{quotation}
Some biological models are known to exhibit mathematical chaos, yet
are considered by biologists to have regular dynamics capable of maintaining
biological function. These systems typically display highly erratic behaviors
at short timescales but maintain regular features at slower, physiologically 
relevant timescales.  We explore this conundrum mathematically and 
identify mathematical structures which allow chaotic deterministic systems 
with multiple timescales to maintain macroscopic regularity. We also 
exhibit a general global bifurcation mechanism that can cause these systems  
to transition to highly erratic behaviors at slow timescales.
\end{quotation}

 \section{Introduction}
\begin{figure*}
	\begin{center}
	\includegraphics[width=.75\textwidth]{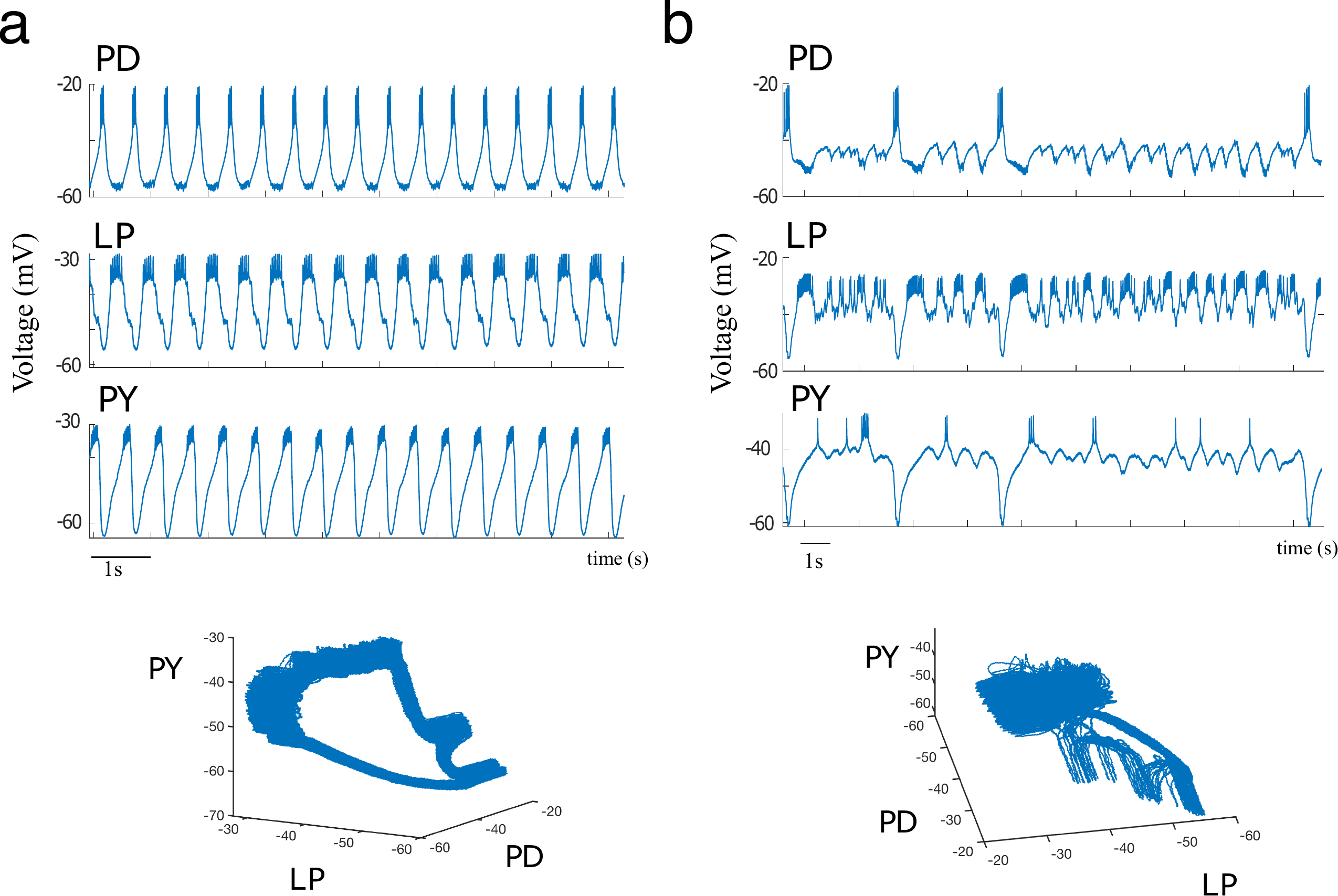}
	\end{center}
	\caption{Intracellular voltage traces from 3 pyloric circuit neurons of the
\emph{Cancer borealis} STG. (a) Chaotic behaviors in an intact system
showing rhythmic bursting with low fluctuations in the period of oscillations 
(with a coefficient of variation $c_v=0.02$), and 
(b) Erratic bursting/firing dynamics in a decentralized network, where
the irregularity in the slow dynamics is characterized by sizable
fluctuations of the relaxation cycle duration ($c_v=0.63$).}
	\label{fig:data}
\end{figure*}
Deterministic dynamical systems that feature chaotic dynamics are
usually characterized by irregular aperiodic patterns of activity
and unpredictable behaviors which are highly sensitive to perturbations 
of parameters and initial conditions. 
These erratic characteristics of chaotic dynamical systems
seem to be undesirable for physiological biological systems that
are required to maintain robust activity patterns to function. However,
the presence of chaotic dynamics has been reported in several
experiments on physiological behavior, 
including key functions such as
cardiac contractions~\cite{hayashi1992chaotic} or brain
activity~\cite{korn}, and many of these systems demonstrate highly
regular dynamics at physiological timescales, with irregular activity 
limited to small timescales and rapid fluctuations. A well-studied biological
system exhibiting such behaviors is the stomatogastric ganglion (STG) of
crustaceans~\cite{marder2} described below. In this paper, we explore how deterministic 
dynamical systems can exhibit this interplay between fast chaos and slow periodicity.  

Beyond its conceptual interest, the interplay between irregular activity
and biological function has implications in
understanding possible consistency between chaos and \emph{homeostasis},
the process by which a variety of functions of living organisms are
regulated in response to changes in the
environment~\cite{bernard1879,may1998glutathione,dijk1999ageing,
aronoff2004glucose}. Various system-level feedback control mechanisms have been 
suggested to maintain macroscopic activity (see e.g., in neuroscience, ~\cite{turrigiano2004homeostatic,
cannon2016synaptic,marder2010coordinating,naude2013effects}). At the
single-cell level and in the absence of an exogenous mechanism,
homeostasis has been suggested to be devoid of chaos~\cite{goldberger1991normal}. 
It was associated with the existence of stable fixed points 
that show little dependence to parameters~\cite{reed2017analysis,golubitsky2018homeostasis,
golubitsky2020infinitesimal} or robust limit cycles~\cite{thomas}. However, an
example of a chaotic system showing considerable robustness is given by the
crustacean STG, a motor neural circuit producing a stable triphasic
bursting pattern of activity that controls the movements of muscles involved
in chewing and filtering food (see review~\cite{marder:07}). This
neural network is remarkably robust to changes in internal and external
conditions (Fig.~\ref{fig:data}a) and is able to maintain constant key
features for functional output such as the order of firing of the neural
populations, relative phases, and duty cycles. 
These macroscopic regularities arise despite a clear
cycle-to-cycle variability typical of chaotic systems (covering a dense
region of the phase space, Fig.~\ref{fig:data}a, bottom) which is also
reflected in various detailed models of the
system~\cite{alonso,alonso2}. However, this chaos can significantly
impact function when the neurohormonal inputs to the STG are blocked 
(a process called \emph{decentralization}), yielding an erratic spiking
behavior with sporadic bursts and a complete loss of rhythmicity,
associated with a loss of function (Fig.~\ref{fig:data}b). 
In this work, we consider the question of whether purely deterministic systems can
display these two types of chaotic behavior and how such systems can switch between
them. While we only briefly address stochastic slow-fast systems, 
the results of this paper are also relevant in this context as well, cf. Fig.~\ref{fig:Stochastic}.

From the mathematical viewpoint, all formal definitions of chaos to date rely on fine
properties of the dynamical system~\cite{sander2015many}, and in particular possible regularities 
arising at slower timescales are immaterial in these characterizations. 
The main contribution of this paper is  to refine the notions of chaos 
in the context of deterministic dynamical systems with multiple timescales so as to
to distinguish whether or not  chaos affects 
the macroscopic behavior of the system at slow timescales. 

This paper proceeds as follows. We start by proposing a definition of slow and fast chaos in section~\ref{sec:slowfast}, before proceeding to a detailed analysis of the 
chaotic Rulkov map in section~\ref{sec:rulkov} with a particular focus on exhibiting these types of dynamics and the transitions between them. Section~\ref{sec:prob} proposes a topological account for the dynamics observed in these two regimes that we validate using a probabilistic model for characterizing the slow-fast behavior of the chaotic Rulkov map. We then explore the generality of these findings by studying a variety of models. We first return to the CPG motivation in section~\ref{section:CPG} and 
introduce a simple network model with behavior similar to the CPG network in Fig.~\ref{fig:data}. 
Section~\ref{sec:universality} goes further into exploring the generality of the behaviors described and introduces two other models which contain slow and fast chaos.
Source codes for all numerical simulations are available at \cite{bib:codes} and the numerical methods used in these simulations are discussed in the Appendix. 

\section{Slow and Fast Chaos }\label{sec:slowfast}

There are several definitions of chaos reflecting the
variety of facets of the phenomenon~\cite{sander2015many}. The most
classical characterizations of chaos involve positive Lyapunov
exponents, topological entropy, or information entropy. 
In our context, a natural question is whether any of these characterizations is able to distinguish 
between chaos on the slow and the fast timescales, 
since from the point of view of applications, these are 
qualitatively quite different. 

Central to our definition of fast or slow chaos, we will focus our
attention on the timing of specific events arising at the slow
timescale. If, for a system displaying chaos, the fluctuations in the
timing of these events are negligible compared to the slow timescale, we
will qualify this chaotic behavior as {\bf fast chaos}. Instead, if the
fluctuations are of the same order of magnitude as the slow timescale,
we will qualify the behavior as \textbf{slow chaos}. 

To make this qualitative definition more quantitative, we propose the
following empirical description. We consider a sequence of events arising 
at the slow timescale For example, in data this time scale could be characterized 
by events such as directional threshold crossings,  
switches between biologically relevant regimes such as the up- and down-states in 
Fig.~\ref{fig:data},  or  Poincar{\'e} sections. The duration of the 
time interval separating such events defines a sequence of real
numbers $(\tau_n)_{n\geq 0}$, with a mean $ \bar{\tau}$ and standard 
deviation $\Sigma$. The so-called coefficient of variation is defined as
$c_v := \bar\tau/\Sigma$. A coefficient of variation $ c_v$ of order $1$ characterizes trajectories where the fluctuations in the timings of the slow events are of the same order of magnitude as their means, which is characteristic of orbits $O_N$ that are slowly chaotic. If instead $c_v$ is small compared to $1$, then the fluctuations in these timings are small compared to the mean, corresponding to trajectories that are regular at slow timescales and thus displaying fast chaos in this parlance. 
Another relevant dimensionless quantity that we will consider is the
level of fluctuations in the timing of the slow events (measured 
using the standard deviation of return times) relative to the slow timescale, $\tilde{\Sigma}=\Sigma\cdot \mu$ where $\mu$ is the inverse characteristic time of the slow variables. If $\tilde{\Sigma}$ is of order one, the fluctuations in the timing of slow events are on the same order of magnitude as the slow timescale, corresponding heuristically to slow chaos. Instead, if $\tilde{\Sigma}$ is small compared to $1$, the events arise at regular intervals compared to the slow timescale and we have fast chaos. 

Maslennikov and Nekorkin~\cite{maslennikov16} 
reported, for the first time to our knowledge, the possibility of the emergence of 
chaotic trajectories with relatively regular slow trajectories in idealized mathematical models
and discrete FitzHugh-Nagumo maps. Such trajectories are akin to what
we classify as fast chaos in this paper. In the same vein, chaotic
trajectories with regular slow dynamics were reported in periodically
forced cubic maps~\cite{han17}. In these systems, the regularity of the
slow dynamics is imposed by a slow periodic variation of a parameter in
the equation, while chaos emerges at a much faster timescale due to the
instabilities of the fast dynamics. Another highly relevant work is the
analysis of Terman~\cite{terman92} of chaotic trajectories in slow-fast
systems of differential equations describing neural systems with
relaxation cycles in the vicinity of a bifurcation from periodic
bursting behavior to spiking behavior. This takes place in the vicinity of homoclinic orbits, 
and may arise in three-dimensional differential equations which do
not include the possibility of fast chaos. 

\section{The Chaotic Rulkov Map}\label{sec:rulkov}
\begin{figure*}[ht]
	\centering
	\includegraphics[width=.75\columnwidth]{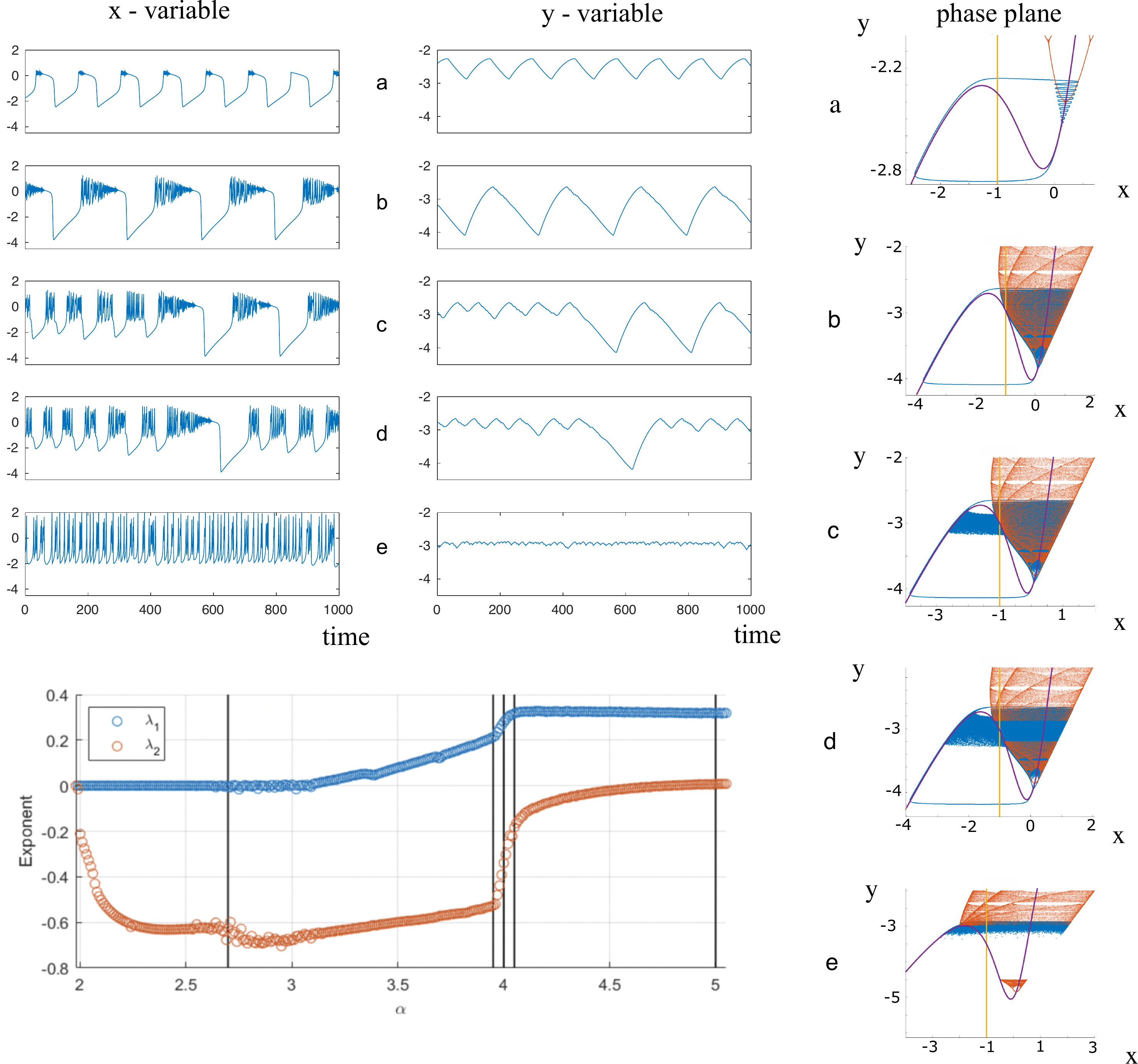} 
	\caption{Top: Chaotic trajectories of the Rulkov map, at parameters $\mu = 0.01, 
	\sigma = -1$ with subfigures  $a,b,c,d,e$ corresponding  resp. to 
	$ \alpha = 2.698$,  $3.95$,  $4.00$,  $4.05$, $5.00$.  
	 Left:~$x_n$; Middle:~$y_n$; Right: Phase planes for $x$ versus $y$ showing an orbit
	 for $\mu=0.01$ (blue), the fast attractor for $\mu = 0$ (red), 
	 the fast nullcline (purple), and the slow nullcline (yellow). 
	 Bottom Left: Lyapunov exponents as a function of $\alpha$, 
	 indicating that chaos emerges when $\alpha \approx 2.6$. 
	   }
	\label{fig:Lyap_Traces}
\end{figure*}
The core of this paper relies on the analysis of the chaotic Rulkov map introduced in~\cite{rulkov2001regularization}. This model was shown to encompass, within a simple
two-dimensional discrete dynamical system preserving mathematical tractability, some of the most prominent neuronal behaviors~\cite{ibarz2011map,courbage2010map}. This map is given by:
\begin{align}
x_{n+1} &= \frac{\alpha}{1+x_n^2} +y_n \\
y_{n+1} &= y_n - \mu \;  ( x_n - \sigma).
\end{align}
The dimensionless variable $x_n$ models the membrane potential of a neuron at discrete time step $n$, and $y_n$ is an adaptation variable accounting for slower gating
processes. The resting potential is denoted by $\sigma$, which will be fixed to $\sigma=-1$ 
for the rest of the study. The small parameter $\mu$ denotes the timescale ratio of
adaptation compared to voltage, and the parameter $\alpha$ controls the
excitability of the neuron. This excitability parameter $\alpha$ will serve as our main bifurcation
parameter in this study. 
As the neuron becomes more excitable, it
 transitions from a resting state to an invariant
cycle~\footnote{This cycle emerges through a Neimark-Sacker bifurcation
arising at $\alpha = -(1-\mu)(1+\sigma^2)^2/(2 \sigma)$ for almost every
$\mu<3$ (for our parameters, $\alpha$ slightly below 2).} which rapidly
grows into a large amplitude relaxation cycle that eventually becomes
chaotic upon increasing excitability, as shown in Fig.~\ref{fig:Lyap_Traces} (see also~\cite[Section 2.1.2]{ibarz2011map}). A fine analysis of these chaotic trajectories identifies distinct classes of behaviors, as we depict in Fig.~\ref{fig:Lyap_Traces}: 

\begin{itemize}

	\item[(a)] For $\alpha$ just above the value associated with the emergence of a positive Lyapunov 
	exponent, we observe marginally chaotic orbits with regular relaxation cycles exhibiting fluctuating
	patterns on the crest of the relaxation cycle. 

	\item[(b)] As $\alpha$ is further increased, we continue to observe relatively regular relaxation cycles, yet bursts become more complex and show higher amplitude spikes with a repetitive triangular profile, which is notwithstanding all distinct and non-periodic. 

	\item[(c)] Beyond a threshold level for $\alpha$, the slow dynamics suddenly becomes irregular, with long bursts with triangular spike profiles (as described in (b)) alternating non-periodically with shorter bursts with approximately constant spike amplitude (yielding rectangular-shaped envelopes).

	\item[(d)] As $\alpha$ is further increased, slow chaos persists, with shorter bursts becoming more frequent and longer bursts rarer.

	\item[(e)] Eventually, the system reaches a threshold value of $\alpha$ beyond which the system no more produces relaxation cycles and the orbit is continuously bursting/spiking, with no discernable separation of bursts. 

\end{itemize}

As expected, the study of Lyapunov exponents precisely identifies the presence 
of chaos. However, these exponents do not appear to identify a transition between 
fast chaos (with relatively regular repetitions of chaotic bursts) and slow chaos with
irregular alternations of longer triangular bursts and briefer square bursts. A close 
inspection of the second Lyapunov exponent shows however an abrupt change in its 
dependence in $\alpha$, with a possible $C^1$ discontinuity, coincident with the switch
between fast and slow chaos. This
can be viewed as a measure of the attraction of points to the chaotic attractor. However, 
this abrupt change may primarily be due to the fact that at the emergence of shorter bursts, 
the decrease in the interburst period results in the system spending less time in quiescence, and
thus there is less time for trajectories to be attracted. All in all, these measures of chaos
just seem to be monotonically increasing with respect to $\alpha$, and
they do not reflect our observation of a loss of macroscopic regularity between regimes 
(b) and (c) and the emergence of an irregular alternation of long triangular bursts and
brief square bursts associated with slow chaos.

To study these dynamics further, we exploit the natural separation of
timescales in the Rulkov map, and separately consider the {\em fast
dynamics} by setting $\mu=0$. This corresponds to considering a 
collection of one-dimensional maps parameterized by the slow adaptation 
variable $y$ now being a constant value. The attractors, depicted in 
red in the phase planes a-e on the right in Fig.~\ref{fig:Lyap_Traces}, show a 
drastic change in structure as $\alpha$ is varied. Typically, the fast dynamics 
feature one stable equilibrium or three equilibria depending on the value of $y$. 
These equilibria lie on the manifold $y=x-\alpha/(1+x^2)$, forming the purple curve in Fig.~\ref{fig:Lyap_Traces} and referred to as the $x$-nullcline. The $y$-nullcline $x=\sigma$
separates regions of the phase plane where the slow variable increases
or decreases, governing the emergence of slow relaxation cycles.

We observe that fast chaos is related to the presence of
relaxation cycles through a chaotic attractor in the fast variable
(cf. Fig. \ref{fig:Lyap_Traces}{b}), as described
in~\cite{maslennikov16}. These regimes affect only the precise behavior
of the voltage during a burst, with little effect on the evolution of
the slow variable, allowing the system to maintain almost
periodic behavior at the slow timescale 
(Fig.~\ref{fig:Lyap_Traces}{b, middle}). In contrast, as $\alpha$ is
increased towards 4 (Fig.~\ref{fig:Lyap_Traces}{c}), the chaotic
attractor intersects with the unstable fixed point of the fast dynamics,
meaning that the system can terminate a burst before completing a full
relaxation cycle. Those {\em shortcuts} appear to be relatively
unpredictable and depend on the particular chaotic sequence arising at
the fast timescale, and they become more frequent as $\alpha$ is
increased. 

In the limit of timescale separation, one would expect the emergence of
slow chaos to correspond to an internal crisis of the fast system;
indeed, for each $\alpha>4$ there are two values of $y$ where such a crisis occurs 
in the 
fast system, where the fast chaotic attractor hits the
unstable fixed point of the fast dynamics. See Section~\ref{section:ProbabilisticSetup} for 
more details. 
For $\alpha>4$, we observe
that the attractor is split into two pieces divided by a region of $y$-values where the
fast dynamics converges to the stable fixed point, cf. the 
fast attractor (in red) and the stable fixed point (in purple) 
in Fig.~\ref{fig:Lyap_Traces}d. 
However, in this non-chaotic region, transient
chaos arises along the ghost of the chaotic attractor, implying that
relatively long transients emerge before the fast system converges
to the fixed point~\cite{grebogi:83,grebogi:85,grebogi:87}. 
These long transients thus compete with the slow evolution of the
adaptation variable. If the duration of the chaotic transient is on the same order 
of magnitude as the time it takes the slow variable to return to the second part 
of the chaotic attractor, a long burst will appear similar to the $\alpha<4$ case. 
Instead, if the duration of the chaotic transient is 
smaller than this time and enters the region of direct convergence to the fast stable
fixed point, then a short burst will emerge. The sequence of long and short bursts 
will thus depend on the chaotic transient and on whether or not the passage through the
ghost of the chaotic attractor goes all the way to the second internal crisis or not.
The system's progression of full cycles and 
shortcuts forms a chaotic sequence, where the frequency with which the system
takes the shortcut varies as a function of $\alpha$ (cf. Fig.~\ref{fig:figure3_histogram}
and Section~\ref{sec:prob} below). 

\section{Statistical analysis of the Rulkov model}\label{sec:prob}

The delicate interplay between chaotic transients and slow dynamics can be further characterized by considering in detail the topology of the attractors and estimating the `likelihood' of a shortcut transition. We finely analyze this in the present section and estimate the frequency with which solutions of the Rulkov system take a shortcut as a function of the parameter $\alpha$, 
as well as looking at how this frequency varies with the choice of $\mu$. 

\subsection{Statistics of the interburst interval}

\begin{figure}[t]
	\centering{
		\includegraphics[width=\linewidth]{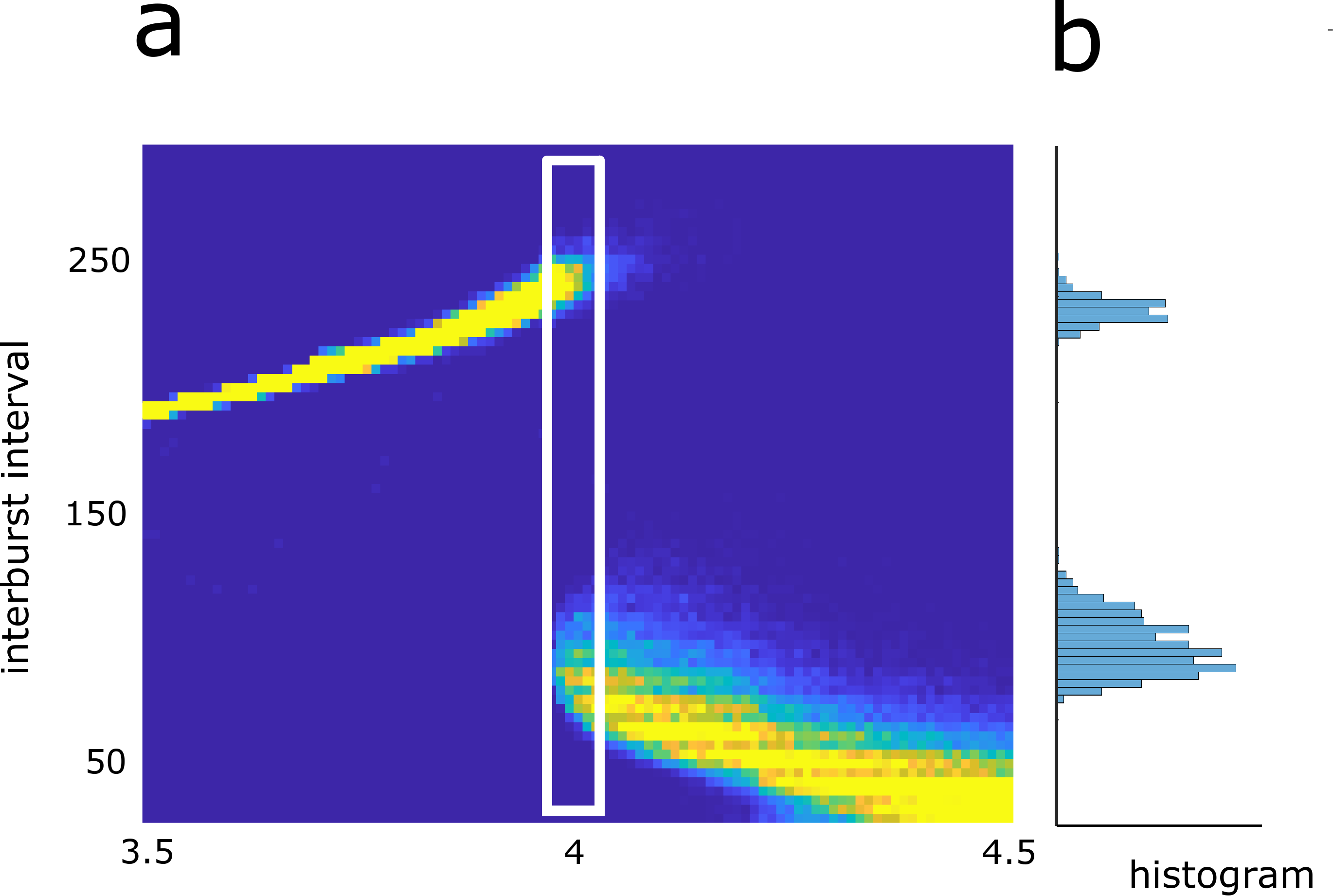}
	}
	\caption{		a. Histogram of the inter-burst interval for $\mu=0.01$ and various
		values of $\alpha$. A bimodal distribution, corresponding with slow
		chaos, arises in a non-trivial range of $\alpha$ in the vicinity of the
		crisis at $\alpha=4$ (white box). b. One typical histogram for 
		$\mu=0.01$ and $\alpha=4$.
		\label{fig:figure3_histogram} }
\end{figure}
To further characterize the fluctuations of the  deterministic dynamics
on the order of the slow time scale, let us consider the sequence of
interburst intervals along a given trajectory, as seen in
Fig.~\ref{fig:figure3_histogram}a. A unimodal, non-Dirac distribution
 corresponds to fast chaos trajectories, with distributions centered
at larger durations for fast chaos with no shortcut (cf.
Fig.\ref{fig:figure3_histogram}a, small $\alpha$), and centered at
smaller durations for fast chaos trajectories where the probability of
shortcut is unity (cf. Fig.\ref{fig:figure3_histogram}a, large
$\alpha$). In contrast, slow chaos corresponds to bimodal distributions
of interburst intervals, corresponding to trajectories composed of both
long bursts and shortcuts (inside the white box in
Fig.~\ref{fig:figure3_histogram}a and in~\ref{fig:figure3_histogram}b). 

\begin{figure*}
	\centering{
		\includegraphics[width=.9\linewidth]{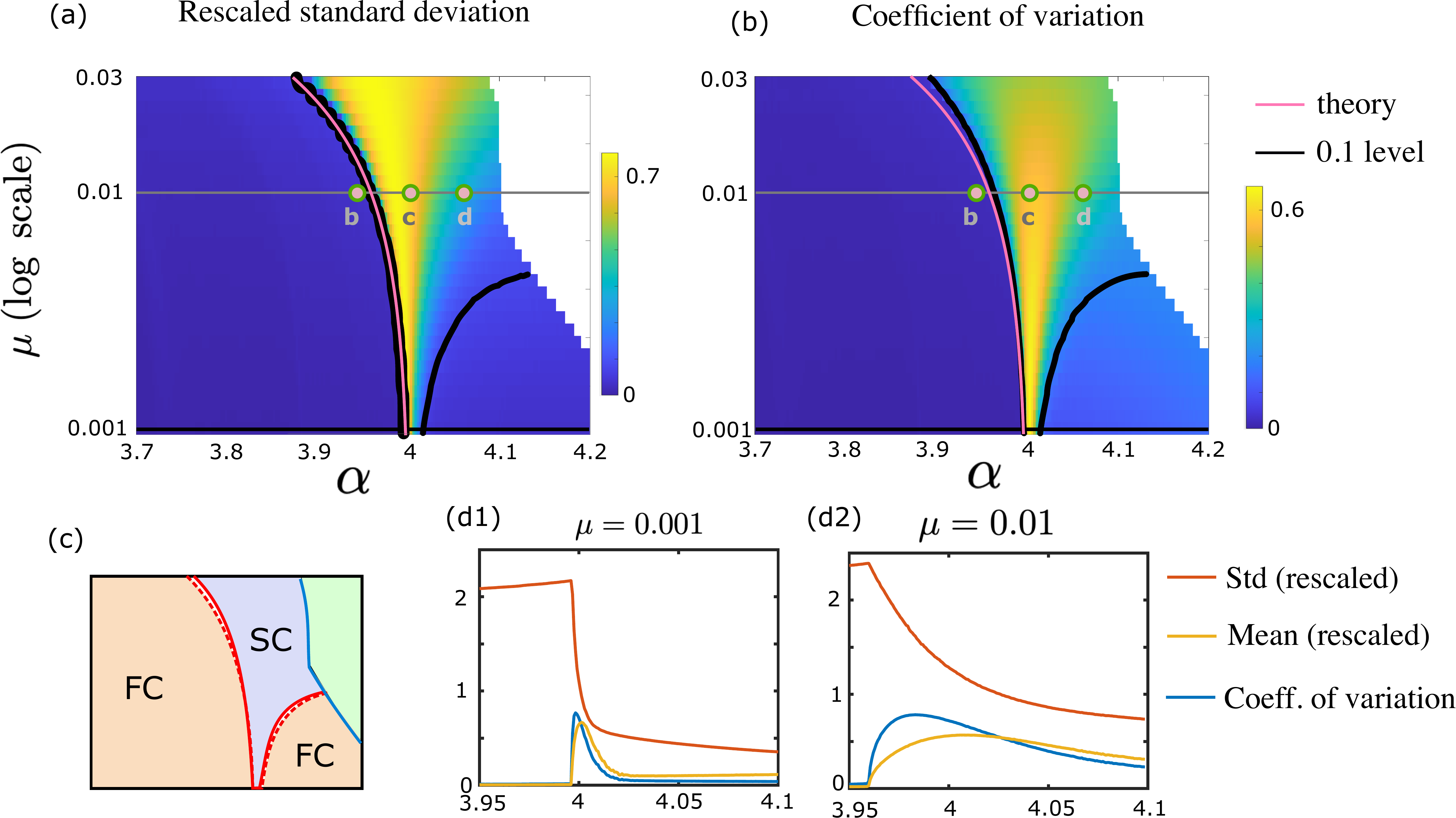}}
	\caption{
	(a) Standard deviation  rescaled by the
	parameter $\mu$. 
	(b) Coefficient of variation $c_v$ 
	of the interburst interval as a function of the parameters $\alpha$ and $\mu$. 
	In both (a) and (b), we observe a sharp transition at the onset of slow chaos. 
	The pink line depicts our theoretical derivation of 
	the onset of slow chaos. The black line depicts the 0.1 level set for the corresponding measure.
	The location of the level sets is consistent with our theoretical prediction. 
	The circles b,c,d 
	indicate the parameter values from resp. Fig.~\ref{fig:Lyap_Traces}b,c,d.
	(c) Bifurcation diagram of the system obtained from (a) and (b): 
	FC: Fast Chaos (beige), 
	SC: Slow Chaos (purple), and
	Loss of Relaxation cycles (green). 
	Bifurcation Lines (red) are the 0.1 levels of rescaled $\boldsymbol{\sigma}$ (dashed) 
	and $c_v$ (solid). 
	(d1,d2) Depictions of the rescaled mean (yellow), rescaled standard deviation (red), 
	and $c_v$ (blue) for $\mu=0.001$ (middle) and $\mu=0.01$ (right).  \label{fig:CoeffVar} }
\end{figure*}

The emergence of slow chaos can be further characterized by either the
coefficient of variation $c_v$ for the interburst intervals (cf. Fig.~\ref{fig:CoeffVar}a) 
or its timescale normalized standard deviation (cf. Fig.~\ref{fig:CoeffVar}b).  
For small $\alpha$, this shows we have fast chaos.
For $\alpha \approx 4$ we have slow chaos, and while the crisis in the fast subsystem occurs 
at $\alpha = 4$, the emergence of slow chaos occurs at an $\mu$-dependent 
$\alpha$ value a bit less than $4$. 
The figure shows that for each $\mu$, the transition to slow chaos is quite rapid, and 
the $0.1$ level set (black line) 
serves as a good choice to define the transition point.   
As $ \mu \to 0$, the $\alpha$-window with slow chaos shrinks to size zero. 
For $\alpha >4$ we have fast chaos again, since orbits reliably take the shortcut. 
Finally, for $ \alpha \gg 4$ the
relaxation cycle breaks down, and we have continuous bursting. 

\subsection{Topological description}
To present a topological description of how slow chaos emerges, we
 consider a coarse-grained representation of the dynamics \cite{gedeon}.  
 Relaxation cycles in the $\mu \ll 1,  \alpha<4$
   regime consist of a concatenation of the following four dynamical sections 
   (see Fig.~\ref{fig:figure3}d): 
\begin{itemize}
	\item[B1-] Slow rise of the adaptation variable preceding a burst.
	\item[T1-] 
	Transition from quiescence to bursting. 
	\item[B2-] 
	Bursting behavior with decay of the slow variable. 
	\item[T2-] Transition from bursting to quiescence.
\end{itemize}
However, for larger values of $\mu$ and $\alpha<4$ sufficiently close to
$4$, we find that the slow dynamics may allow a return to the left
stable fixed point through the emergence of short bursts (marked as T3 in Fig.~\ref{fig:figure3}d).

For $\alpha>4$, because of the presence of two $y$-values with crises, 
shortcuts exist for arbitrarily low values of $\mu$ between the 
two  crisis $y$-values. In fact, shortcuts become the most likely outcome as $\mu\to 0$, 
and therefore coarse-grained transition graph must be slightly modified (Fig~\ref{fig:figure3}e). 
In that case, the transition
region from bursting to quiescence is split into two disconnected
pieces B2 and B3, with returns to quiescence denoted by T2 and T3. For $\mu$
sufficiently small, fast chaos occurs with relaxation cycle  B1-T1-B2-T3, 
but when $\mu>0$, the transient chaotic behavior may prevent the
system from undergoing transition T3; in this case, cycles transition to B3 through
a new path B4, yielding a long bursting relaxation cycle
 B1-T1-B2-B4-B3-T2. 

\subsection{Probabilistic model}\label{section:ProbabilisticSetup}

\begin{figure*}
	\centering{
		\includegraphics[width=\linewidth]{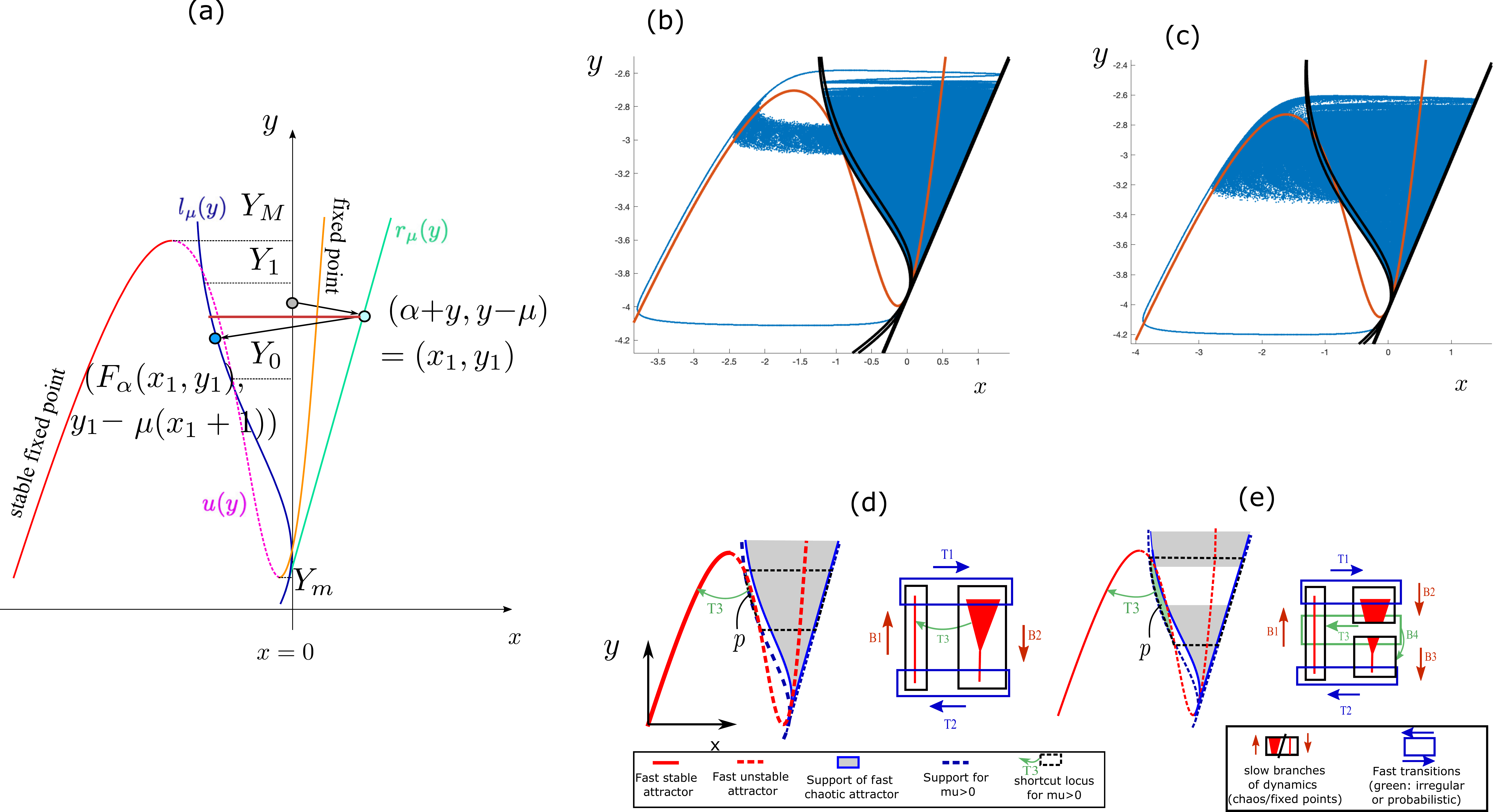}}
	\caption{Probabilistic model and topological dynamics. 
	(a) The dynamical
structures and our approximation of the support of the fast attractor.
This figure highlights the main notation used for the probabilistic
model, in particular the left and right envelopes $l_\mu(y)$ and
$r_\mu(u)$, the unstable fast equilibrium $u(y)$ and the possible
intersections of the $l_{\mu}$ manifold and the unstable fast
equilibrium manifold $Y_0$ and $Y_1$. 
The red horizontal line from $l_\mu(y)$ to $r_\mu(y)$ 
is $I_{\mu(y)}$. 
Panels (b) and (c) illustrate the
slow chaos dynamics in the case of $\alpha<4$ (b) or $\alpha>4$ (c,
specifically, $\alpha=3.98$ and $\alpha=4.02$, represented for
$\mu=0.01$). Blue: trajectory, showing an excellent fit with the envelope computed analytically (black, outer curve; the inner black curve corresponds to $\mu=0$), Ref: fixed points of the fast system. 
(d-e) Cartoons of the dynamics and topological behavior. (d) slow chaos for $\alpha<4$ (here, $\alpha=3.98$ and $\mu=0.02$), where the support of the chaotic attractor
does not intersect the unstable fast equilibrium but the slow variable
may allow the system to take shortcuts $T_3$. 
(e) slow chaos for $\alpha>4$ (here, $\alpha=4.02$ and $\mu=0.02$) where shortcuts arise for arbitrarily small values of $\mu$, but for $\mu>0$ chaotic transients may allow jumping
between the two parts of the chaotic attractor (transition $B_4$). }
	\label{fig:figure3}
\end{figure*}

Based on these observations, we have designed a probabilistic model to
predict the emergence of slow chaos and estimate the shortcut
probability (see Fig \ref{fig:figure3}a). This model simply follows the 
Rulkov dynamics in the region of the phase plane where we expect non-chaotic
branches of relaxation cycles  (i.e., to the left of the unstable fixed point marked in yellow
on the graph),
and it randomly samples the support of the chaotic attractor otherwise. 

More formally, our stochastic model approximates the fast chaotic dynamics with 
a random Bernoulli sequence (independent random variables) uniformly sampling
the interval of possible values for the fast variable along a relaxation cycle. 
In particular,  let $[Y_0,Y_1]$ be the region of (slow variable)
 phase space where the transition $T3$ is accessible (cf. Fig.~\ref{fig:figure3}a),
 let $\mathcal{I}_\mu(y)$ be the interval of fast variable values  
during the high voltage portion of the burst, i.e. for which the fast dynamics are chaotic 
(red horizontal line in Fig.~\ref{fig:figure3}a),  
 and define $\mathbb{U}_I$ to be a uniform random variable on this interval.  
Our model is given by: 
\[\begin{cases}
 x_{n+1}=\begin{cases}
 	\mathbb{U}_{\mathcal{I}_\mu(y_n)} & Y_0<y_n<Y_1 \; \text{and}\; x_n>u(y_n)\\
	F_\alpha(x_n,y_n) & \text{otherwise}\\
	\end{cases}\\
 y_{n+1}=y_n - \mu(x_n-\sigma) \; . 
\end{cases}\]
Here,  $F_\alpha(x,y)= \frac{\alpha}{1+x^2}+y$ is the nonlinear fast dynamics,   
and $u(y)$ is the unstable
fixed point of the fast dynamics.

Note that while an exact representation $\mathcal{I}_\mu(y)$ is only given implicitly, noting 
that the fast variable takes its maximum at $x=0$, this envelope is approximated by the first and 
second iterates of $x=0$, yielding a closed-form expression that closely matches the numerically 
observed support of the fast variable (see Fig.~\ref{fig:figure3} b,c, black curves). The 
significance of this representation is that it provides
a quantification of the relative occurrence of a shortcut between the two $y$ values 
associated with crisis points, where the 
deterministic chaotic system provides no information. 

We now return in more detail to the slow-fast decomposition of the dynamics, 
using  the dynamics of the fast variable $x$ (fast dynamics) when the slow variable $y$
is frozen. For each fixed $y$, this one-dimensional dynamical system is given by the
function 
\[x \mapsto F_{\alpha}(x,y)=\frac{\alpha}{1+x^2}+y.\] 
The fixed points of
this equation are thus the roots of the cubic polynomial: 
\[x^3-x^2 y +x - y -\alpha=0.\] 
Therefore, depending on $y$ and $\alpha$, the system has between 1
and 3 solutions, corresponding to intersections of and fixed values of $y$ and 
the curve $y = x-\frac{\alpha}{1+x^2}$, forming the $x$-nullcline for the original map. 
In the case where there are three solutions, the middle fixed point is
necessarily unstable and we denote it by $u(y)$, for a given $y$ and
fixed $ \alpha$ (cf. Fig. \ref{fig:figure3}).

 The fast dynamics $x \mapsto F_{\alpha}(x,y)$ is
	a unimodal map, and we observe numerically that increasing $\alpha$ leads to a period-doubling route to chaos, a classical result of a class
	of unimodal maps~\cite{alligood-yorke}. 
To compute a region enclosing the chaotic attractor $\mathcal{I}_\mu(y)=\{x: l_\mu(y)<x<r_\mu(y)\}$, 
for each $y$, 
we estimate an upper-bound using the maximal value of
$F_\alpha(\cdot,y)$, which is attained at $x=0$. Thus the graph of $r_\mu(y)$ is given by 
$\{(x_1,y_1)  = (F_\alpha(0,y),y-\mu)=(\alpha+y,y-\mu): y \in \mathbb{R} \}$. 
The left side of the interval, meaning the 
graph of $l_\mu(y)$, is given by the image of the points $(x_1,y_1)$ on the graph of $r_\mu(y)$. 
That is, 
the graph of $l_\mu$ is given by 
\begin{eqnarray}
	\{ &(&\l_\mu(y_2),y_2) = (F_\alpha(x_1,y_1),y_1-\mu(x_1+1)) \nonumber \\
	& =&(F^2_\alpha(0,y),y_1-\mu(x_1+1)) \nonumber \\
	&=& \left( \frac{\alpha}{1+(\alpha+y)^2}+y,y-\mu -\mu(\alpha+y +1) \right) :\nonumber \\ 
	&y& \in \mathbb{R}  \} \, ,
\end{eqnarray}
cf. the solid blue and green lines in Fig~\ref{fig:figure3}a, and 
show an excellent agreement with the simulations in Fig.~\ref{fig:figure3}d,e.

We know that the sequences of
iterates will be contained in this interval as long as
$F_\alpha(\mathcal{I}_\mu(y),y))\subset \mathcal{I}_\mu(y_1)$, which is in particular the case
when $l_\mu(y)$ is to the left of $u(y)$. Moreover, if
$x=0$ belongs to the support of a chaotic attractor, then this support
contains the boundaries of the interval $\mathcal{I}_\mu(y)$. In this
situation, an internal crisis will arise for parameters and a slow
variable $y$ such that $l_\mu(y)$ falls exactly on $u(y)$. 
The two points at which this occurs are $Y_0$ and $Y_1$ depicted in Fig.~\ref{fig:figure3}a. 
In the 
case $\mu = 0$, the condition yields
\begin{displaymath}
	2 y^2 + 3\alpha y+ \alpha^2+2=0 \; . 
\end{displaymath} 
and the roots of the equation are given 
$\frac{1}{4}(-3\alpha\pm \sqrt{\alpha^2-16})$, 
with two real roots  when $\alpha > 4$ (and a single real root at $\alpha=4$). 
Thus the chaotic attractor is contained 
in an interval tangent to $u(y)$. Two real solutions correspond 
to the two internal crises delineating the
region of occurrence of slow chaos within $y \in [Y_0,Y_1]$. 
Note that if $\alpha > 4$ and $\mu > 0$, but sufficiently 
small, then there will  still be  two ($\mu$-dependent) points $Y_0$ and $Y_1$  
of intersection between $l_\mu(y)$ and $u(y)$. We will see below that in fact
for small fixed $\mu>0$, the interval $[Y_0,Y_1]$ persists for a small region of $\alpha<4$.

\subsection{A uniform random variable on the support}
\label{subsec:RandomVariableOnSupport}
In the model introduced in the the previous section, 
for $y \in [Y_0,Y_1]$, we have modeled the system  such that it uniformly 
samples the region $\mathcal{I}_\mu(y)$ as 
it moves down along the slow flow, where we consider 
the fraction of length of the interval $\mathcal{I}_\mu(y)$ to 
the left of $u(y)$. This provides an approximate probability for each the 
iterate to lead to a shortcut trajectory. The probability of a long burst is thus 
given by the probability of never falling to the left of $u(y)$, which can be readily quantified. 

In our stochastic model, the number of steps performed within the region $y \in [Y_0,Y_1]$ varies
inversely proportionally to $\mu$. For $\alpha>4$, since the interval $[Y_0,Y_1]$ always contains the set of values of slow variables for which $\mathcal{I}_0(y)$ is to the left of $u(y)$, the probability of a short burst increases as $\mu$ is decreased. In contrast, for $\alpha< 4$,
decreasing $\mu$ enough will eventually prevent any intersection between
$\mathcal{I}_{\mu}(y)$ and $u(y)$ for all $y$, because of the continuous dependence
upon parameters and the fact that there is no intersection with
$\mathcal{I}_0(y)$. 

\begin{figure}
	\centering{
		\includegraphics[width=.8\linewidth]{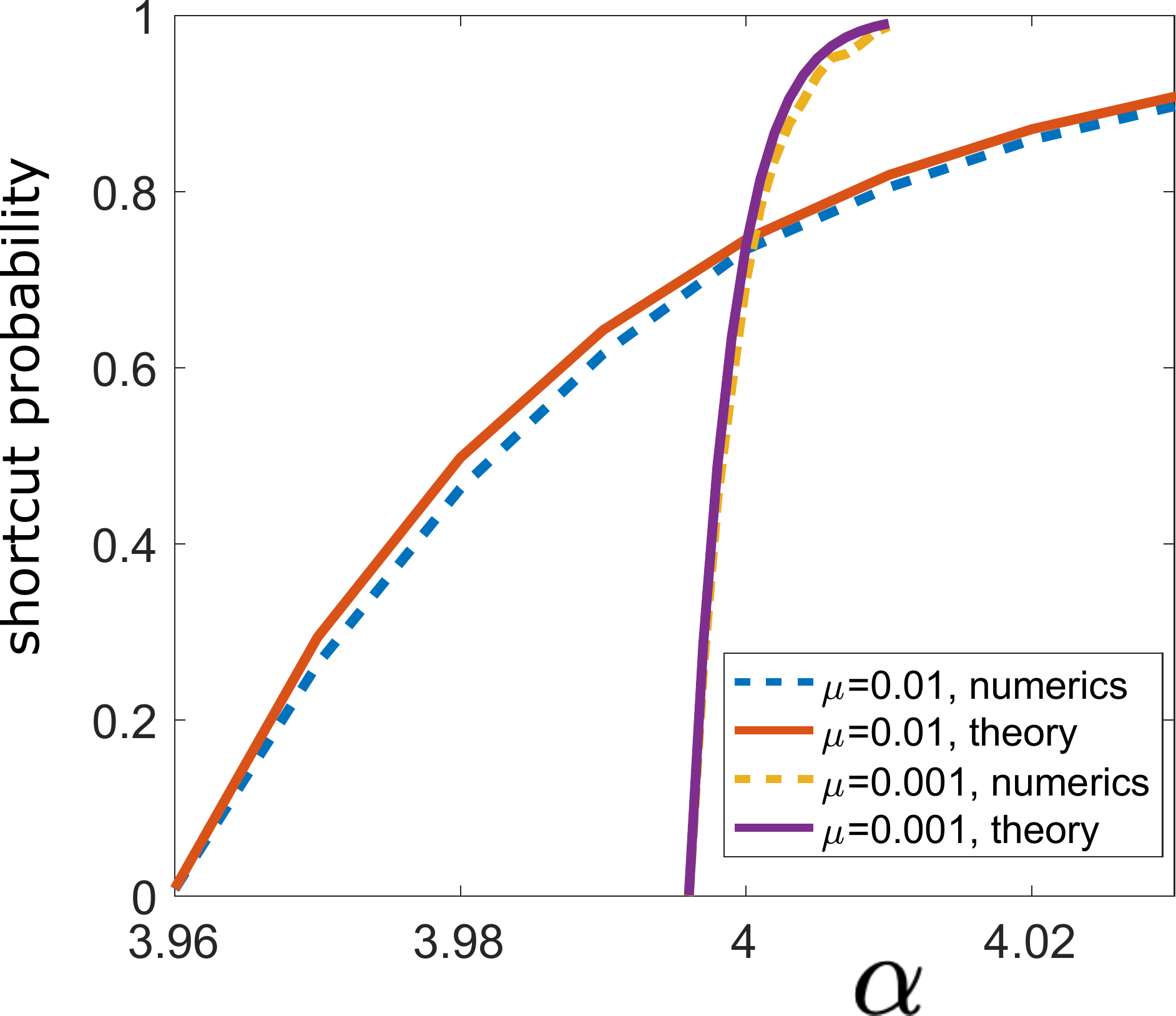}
	}
	\caption{Slow chaos and probabilistic model.  The shortcut probability simulated shows
		an excellent agreement with our analytical expression for the
		probabilistic model. 
		\label{fig:figure3_Switching} }
\end{figure}

We thus expect that, if $\mu$ is sufficiently close to 0, the probability of
a short burst will suddenly jump from 0 and 1 at $\alpha=4$. This is
visible in Fig.~\ref{fig:figure3_Switching}, where we show the joint dependence of
the probability of a short burst in $\alpha$ and $\mu$ (computed with numerical integration). 

Let us now use this probabilistic model to estimate the alternation of shortcuts and long bursts for a fixed value of $\mu$. If the attractor never intersects the unstable 
fixed point (i.e., $l_\mu(y)>u(y)$ for all $y$ where both functions are defined), long bursts
arise systematically, and the system displays fast chaos. However, if there exists an interval 
of values $y \in [Y_0,Y_1]$ such that $l_\mu(y)<u(y)$, an iterate of the stochastic system
leads to a shortcut with ``instantaneous'' probability $\rho_\mu(y)=(u(y)-l_\mu(y))/(r_\mu(y)-l_\mu(y))$. 
The probability of not taking the shortcut per burst will thus be given by the
product of $\prod_{n=n_0}^{n_1} (1-\rho_\mu(y_n))$ where $n_0$ is the time
$y_n$ enters $[Y_0,Y_1]$, and $n_1$ is the exit time. 

While this can be
computed explicitly, we further simplify the problem by considering an
averaged shortcut probability computed as the fraction of surface area $p$
to the right of the unstable fixed point (green region in Fig.~\ref{fig:figure3}d,e), 
and the number of iterates is estimated as
$N=(Y_1-Y_0)/\mu$, the probability of a long burst is $P_l=(1-p)^{N}$.
Numerical calculation of $p$ under this assumption shows very good
agreement with the original simulations of the  Rulkov model 
as we show in Fig.~\ref{fig:figure3_Switching}. Going
further, we observe that the exponential dependence in $1/\mu$ makes the
probability switch from $0$ to $1$ very rapidly upon variation of the
time scale parameter $\mu$ at a given value of $\alpha$ (i.e., for $p$
fixed). This indicates that in the limit of timescale separation, the
system always displays fast chaos, with long bursts for $\alpha<4$
and short bursts  for $\alpha>4$. As $\mu$ is decreased, slow chaos occurs
on a smaller interval of $\alpha$, as
shown in  Fig. \ref{fig:CoeffVar}.
This approach provides a very accurate value, for $\mu>0$, of the emergence of 
slow chaotic trajectories for typical $\alpha<4$, as seen 
by comparing the solid and dashed lines in Fig.~\ref{fig:figure3_Switching}. For any fixed value of $\alpha$, there
exists $\mu$ sufficiently small for which the system displays fast chaos only, 
characterized by relaxation oscillations. 

The fine analysis above is made possible by the simplicity of the Rulkov
map, but it also reveals a robust phenomenon: relaxation cycles through fast
chaotic attractors will yield mathematically chaotic behaviors that
appear regular at larger timescales, and these phenomena will breakdown
at internal crises of chaotic attractors where chaotic alternations of
longer and shorter cycles will emerge. To provide further evidence that this occurs 
for fast slow systems with attractor crises,
subsequent sections demonstrate that the same 
behavior occurs in three other models: a network model of the 
Rulkov map with the topology of the crab STG (Sec.~\ref{section:CPG}), a cubic discrete dynamical
showing a double crisis (Sec.~\ref{sec:cubicmap}); and  a 5-dimensional FitzHugh-Nagumo neuron model 
coupled with the Lorenz system at a fast timescale (Sec.~\ref{sec:FitzHugh-Nagumo-Lorenz}). 

\section{Network of Rulkov maps and Central Pattern Generators}\label{section:CPG}
\begin{figure*}
\includegraphics[width=.9\textwidth]{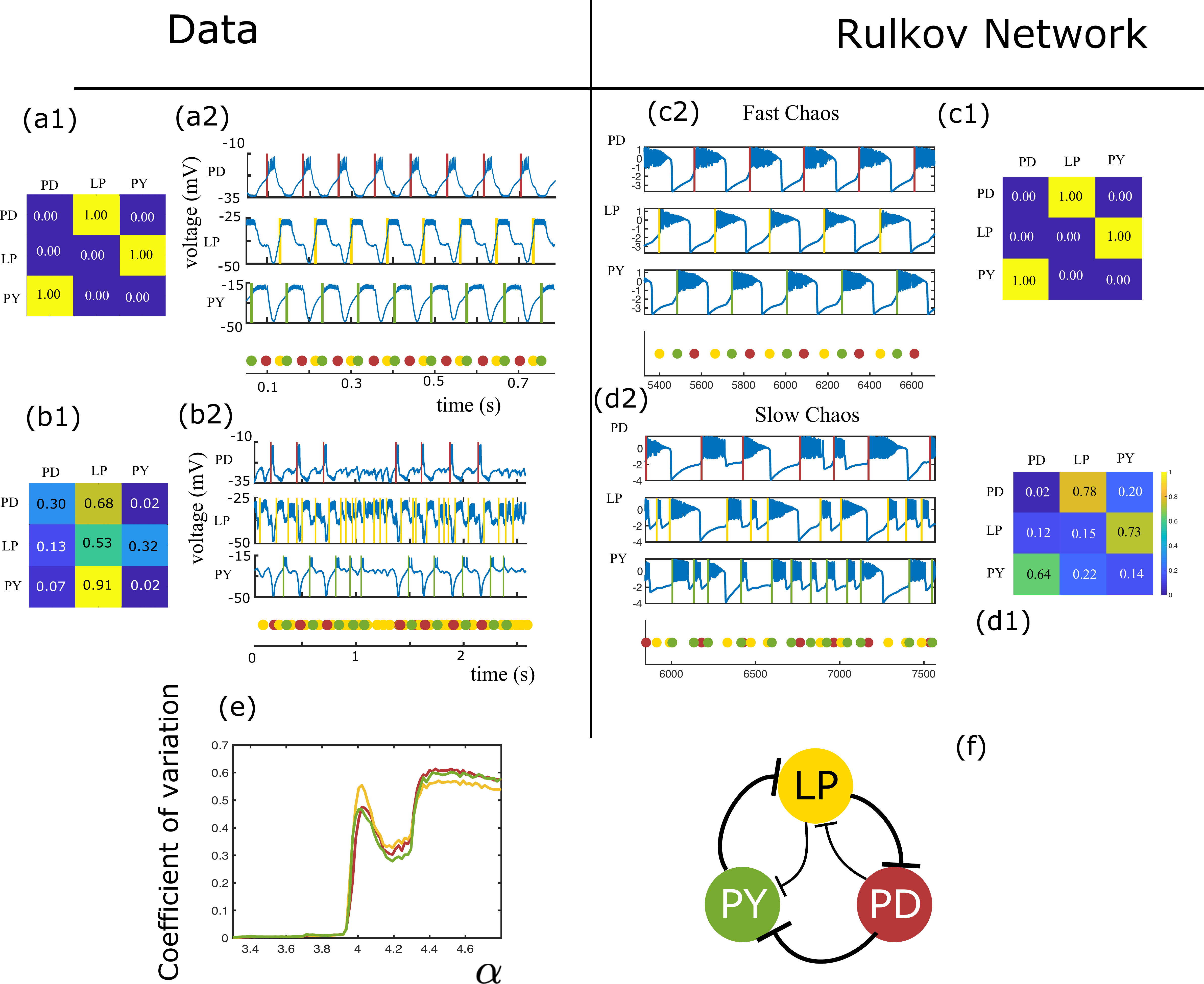}
\caption{
\label{fig:network}
Left: experimental data from~\cite{marder2}. Right: simulations of a 
three-neuron Rulkov model with inhibitory connectivity mimicking the 
pyloric network, with architecture depicted in panel (f). (a) Intact network, (b) Decentralized biological 
network with irregular activity, (c) Rulkov model in the fast chaos regime 
($\mu=0.01$, $\alpha=3.8$) and (d) Rulkov model in the fast chaos regime. 
($\mu=0.01$, $\alpha=4$). (a2,b2,c2,d2): Voltage traces
for each neuron with, on the bottom, a stop-light representation of the spike 
sequence. The associated transition matrices are shown in (a1,b1,c1,d1). The coefficients
of variations for the depicted trajectories are: (a2) 0.01, (b2) 0.60, (c2) 0.02, (d2) 0.62, and the evolution of the coefficient of variation for each population (same color code), depicted in panel (e), shows a clear transition slightly before $\alpha=4$  from low levels associated with fast chaos to large levels associated with slow chaos. }
\end{figure*}
We next explore the dynamics of a simple phenomenological model of the crab STG made of three identical chaotic Rulkov neurons with a biologically-inspired connectivity map (see
Fig.~\ref{fig:network}, right). 
Specifically, the connectivity parameters are 
chosen randomly as 
\[\left(\begin{array}{ccc} 0 & -5.5 &-2.5\\-3.5&0 &
-5\\ -3&0&0 \end{array}\right)+U \, ,\] 
where  $U$ is a random matrix with off-diagonal elements chosen from a 
uniform distribution between $[-0.5,0.5]$, and zeros on the diagonal.
Of course, the neural network of the central pattern generators of the crab
discussed in the introduction involve neurons with vastly richer
dynamics than the Rulkov map and with heterogeneous properties. 
The present model allows us to test the relevance of the transition to slow chaos in a
network without complications associated with intrinsic dynamics or
heterogeneities, although this would be a simple extension of our
analysis. 

We compared our model in the regime of fast chaos and slow chaos to two
typical sequences recorded experimentally in the crab STG (data from
Kedia and Marder~\cite{marder2}) that qualitatively corresponded to 
fast chaos (intact network Fig.~\ref{fig:network}a) or slow chaos 
(decentralized network Fig.~\ref{fig:network}(b). We simulated our network and tested for the sequence of
neurons spiking, deemed to be biologically relevant information. For
that purpose, we identified the onset of each burst and extracted the
sequence of burst initiations (colored circles below the neural dynamics
traces) identified using a threshold-crossing condition
(thresholds were adjusted according to the experimental traces for each
neuronal type, and fixed to $-1.5$ for the model), which led to the stop
lights dynamics represented below the traces. This sequence of spikes 
was used to construct  a ``transition probability matrix'' $M$ that
contains as element $M_{ij}$ the empirical frequency that neuron $j$
spikes after neuron $i$. This is a stochastic matrix, since all elements
are non-negative and the sum over all columns on a given line is equal
to $1$. Given these matrices, we further computed an \emph{entropy}
level, computed as the average entropy of each line, or $-\frac {1}{3}
\sum_{i,j} M_{ij} \log(M_{ij})$ with the convention that $x\log(x)=0$
for $x=0$. A deterministic sequence gives an entropy equal to $0$, while
a maximal entropy of $\log(3)\approx 1.0986$ corresponds to a completely
random sequence of spikes.  

We compared our model in the regime of fast
chaos and slow chaos to two typical sequences that correspond to fast
chaos vs slow chaos recorded experimentally~\cite{marder2}. We observed
that both cases of fast chaos are certainly regular enough to ensure a
fully predictable sequence of spikes with probability 1 for the next
spike neural population to spike, and thus entropy zero. However, the
typical trace of a slow chaos experimental data shows a very irregular
sequence, with evidence of the presence of shortcuts (shorts bursts)
very evocative of those observed in the Rulkov model (see left column,
bottom). This is further shown in the coefficient of variation of the network
as a function of $\alpha$, showing a sudden jump from small fluctuations
to larger values arising again in the vicinity of $\alpha=4$ where we  showed 
the emergence of a crisis in the Rulkov map. 
The experimental dynamics (left) are quite comparable to the
network model we designed, and in
particular the transition matrix in this regime is non-trivial, with
strictly positive probabilities to have any neuron spiking after any
other neuron, non-trivial entropy (0.67 for the data, 0.75 for the
model) and larger coefficients of variations of interburst intervals (see Fig.~\ref{fig:network}(e)). 

\section{Crises yield transitions between slow and fast chaos in other chaotic slow-fast models}\label{sec:universality}
To further explore the generality of our observation of dynamical structures underlying fast chaos and their transitions to slow chaos, we explore two other examples of multiple timescales dynamical systems, illustrating that our results are indeed not specific to a single system, and rather occur 
in a variety of slow-fast systems with crises, both in discrete-time maps and differential equations. 

\subsection{Cubic map}\label{sec:cubicmap}
\begin{figure*}
\centerline{\includegraphics[width=.8\textwidth]{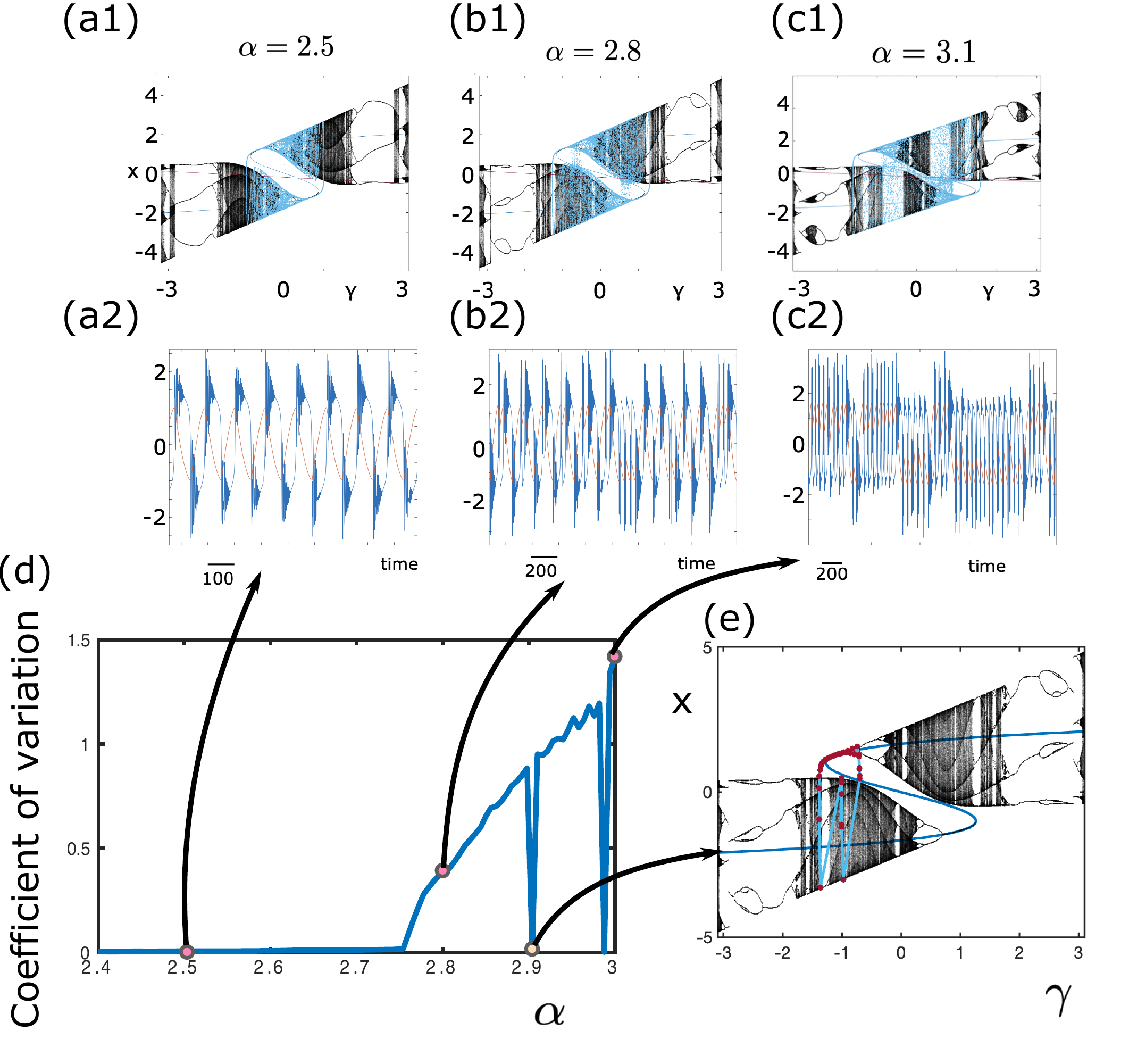}}
	\caption{Cubic map simulations for $\alpha\in\{2.5,2.8,3.2\}$,
	illustrating respectively fast chaos and a transition to slow chaos
	as the system approaches and crosses the crisis in the fast system
	(arising at $\alpha=3$). Top: phase space, with the fast attractor
	for various values of $\gamma$ (black) as well as the trajectory of
	the system for $\varepsilon=0.01$ (cyan). Middle: evolution of the
	fast variable $x$ (blue) and slow variable $\gamma$ (orange) as a
	function of the iterate number, illustrating clearly the emergence of
	fast chaos whereby macroscopically regular but microscopically
	chaotic behaviors emerge, and their transition to slow chaos with
	irregular dynamics when shortcuts in the relaxation cycles can arise
	during transient chaos near crises. Bottom: the coefficient of
	variation of the system as a function of $\alpha$ shows a sudden
	increase at the emergence of slow chaos, with sudden drops
	associated with the emergence of periodic orbits (bottom right, for
	$\alpha=2.904$). }
	\label{fig:cubic}
\end{figure*}
In~\cite{han17}, a one-dimensional, three-parameter cubic map was introduced and shown to feature complex dynamics. This system is given by the equation:
\[x_{n+1}=(-\beta x_n^3 +\alpha x_n+ \gamma) \, .\]
It has been shown to display chaos for specific values of the parameter
$\gamma$ with or without internal crises depending on the value of
$\alpha$. We  exploit this dynamical richness to explore the
existence of fast and slow chaos and the transitions between the two
regimes. To this purpose, we extend the system to allow slow
autonomous variation of $\gamma$, with linear dynamics according to $x$,
in such a way that the system will tend to produce relaxation cycles
through the chaotic attractors:
\[
\begin{cases}
x_{n+1}&=(-\beta x_n^3 +\alpha x_n+ \gamma_n)\phi(x_n)\\
\gamma_{n+1}&= \gamma_n+\varepsilon(-\kappa \gamma_n-x_n+\delta-\mu x_n^3).
\end{cases}
\]
In the equations above, we added for stability of the system a cubic
term $-\mu x^3$ to the linear dynamics of $\gamma$ and a multiplicative
term $\phi(x_n)$ to the $x$-dynamics preventing divergences when $x$ is
large. Here, $\phi$ is a smooth function with a plateau equal to 1 and 
decaying to zero as $\vert x\vert \to \infty$~\footnote{Here, we chose
\[-\frac 1 2 (\tanh(x-\theta)+\tanh(-x-\theta))\] with $\theta=2$.}.
Parameters are fixed to $\beta=0.5$, $\kappa=0.1$, $\delta=0.2$, and $\mu=0.9$.
(These parameters were chosen arbitrarily to ensure the creation of appropriate 
relaxation cycles given the cubic dynamics.) The timescale of the evolution of $\gamma$ is governed 
by the parameter $\varepsilon$ and has been set to $0.01$ in our figure. For these parameters, the
fast variable typically displays chaos for appropriate values of
$\gamma$. The fast variable features hysteresis between two stable fixed
points for $1<\alpha<2$, which is associated with standard relaxation cycles
for the full system. 

Chaos emerges on both branches at $\alpha=2$, and initially,
these fast chaotic trajectories are confined above or below the unstable fixed point.
In that configuration, consistent with our analysis of the Rulkov system, we 
observe fast chaos, with regular relaxation cycles that include a passage through 
a chaotic attractor (Fig.~\ref{fig:cubic}a1) and  hysteresis between
chaotic attractors starting from $\alpha=2$, associated with fast chaos
trajectories with a relatively regular slow dynamics (red curve, Fig.~\ref{fig:cubic}a2)
and fast trajectories that show chaos at a fast timescale but almost
periodic behavior at the slow timescale. To identify the duration of each relaxation
cycle, we used a threshold-crossing condition. We computed, for each value of the 
parameters, the location of the saddle-node bifurcations of the fast variable. We 
chose the leftmost fold and identified one relaxation cycle by considering times when
the fast variable when from behind larger to smaller than this value and with the slow variable 
being below the value of $\gamma$ associated. For $\alpha=2.5$ as in Fig.~\ref{fig:cubic}a, we found a coefficient of variation equal to $0.008$, indicating that the period of relaxation cycles fluctuates by less than 1\%, corresponding to a clear instance of fast chaos. 

As $\alpha$ is increased, the chaotic attractors of the fast system approach the unstable fixed point, until both of them hit the unstable fixed point curve tangentially for $\alpha=3$ at a double-crisis (the coincidence of the two crises is associated with the symmetry of the system). This bifurcation determines the breakdown of fast chaos and the emergence of slow-chaos trajectories. For the full slow-fast system, similar to the case of the Rulkov map, we observe that slow chaos emerges a little before the value of $\alpha=3$ due to the dynamics of the slow variable (according to Fig.~\ref{fig:cubic}d, at around $\alpha=2.76$, and this value shall approach $3$ as $\varepsilon$ is decreased). At $\alpha=2.8$ we clearly observe slow chaos with irregular slow dynamics (red curve in Fig.~\ref{fig:cubic}b2, coefficient of variation $0.311$) associated with whether or not, during the chaotic transient associated with the region of values of $\gamma$ between the two crises, the trajectory crosses the unstable fixed point short-cutting the relaxation or performs a long relaxation cycle, which persists as $\alpha$ is further increased (for $\alpha=3$, Fig.~\ref{fig:cubic}c1-c2 show the presence of slow chaos with a coefficient of variation $1.184$). 

 We note that our systematic computation of the coefficient of variation as a function of $\alpha$ also included sudden drops to a standard deviation of zero. 
 We selected one of these points (here, $\alpha = 2.904$, Fig.~\ref{fig:cubic}e) and observed that for this particular parameter value, the system converged towards a periodic orbit of period 37 instead of performing relaxation cycles. 

\subsection{FitzHugh-Nagumo-Lorenz model}
\label{sec:FitzHugh-Nagumo-Lorenz}

A natural question that could arise at this stage is whether the
alternation of short and long bursts in the Rulkov map or the cubic
map could be associated with the discrete nature of the dynamics, and
would disappear in differential equations. To 
explore this question numerically, we designed a differential equation
that couples relaxation cycles and chaos. 
The slow-fast systems discussed above
are both non-invertible two-dimensional maps. Thus 
to see the same behavior in a differential equation, 
one would expect to need at least a four-dimensional model. 
(In fact, the  model we constructed 
is five-dimensional, but we expect four dimensions would suffice.) 
We constructed our system by
coupling two classical models: the FitzHugh-Nagumo neuron
model~\cite{fitzhugh1,fitzhugh2,nagumo} well-known to produce relaxation
cycles (variables $(r,h)$ in our equations); and the celebrated Lorenz
system, well known to produce chaotic trajectories~\cite{lorenz}
(variables $(x,y,z)$). We coupled these systems by using the slow
variable of the FitzHugh-Nagumo system as a bifurcation parameter of the
Lorenz system. In that way, as the FitzHugh-Nagumo variables perform a
relaxation cycle, the Lorenz system will go in and out of its chaotic
regime. We used a single Lorenz variable ($x$) as an input
to the FitzHugh-Nagumo fast variable, rescaled by a coefficient
$\delta$. For small $\delta$, the chaotic
fluctuations of the Lorenz equation will have a smaller impact on the
FitzHugh-Nagumo dynamics, while for a large $\delta$, the impact will be
more substantial and may break down the relaxation cycles. The resulting 
system of equations is given by:
\begin{equation}
\begin{cases}
	\dot{x} &= \sigma \,(y-x)\\
	\dot{y} &=-x\,z+\gamma \,(\nu+r)\, (\nu+x)-y\\
	\dot{z} &= x\,y-b\,z\\
	\\
	\dot{h}&=h-h^3/3-r+ \delta x\\
	\dot{r}&=\varepsilon\,(h-r) \, .
\end{cases}
\end{equation}

\begin{figure*}
\centerline{	\includegraphics[width=.8\textwidth]{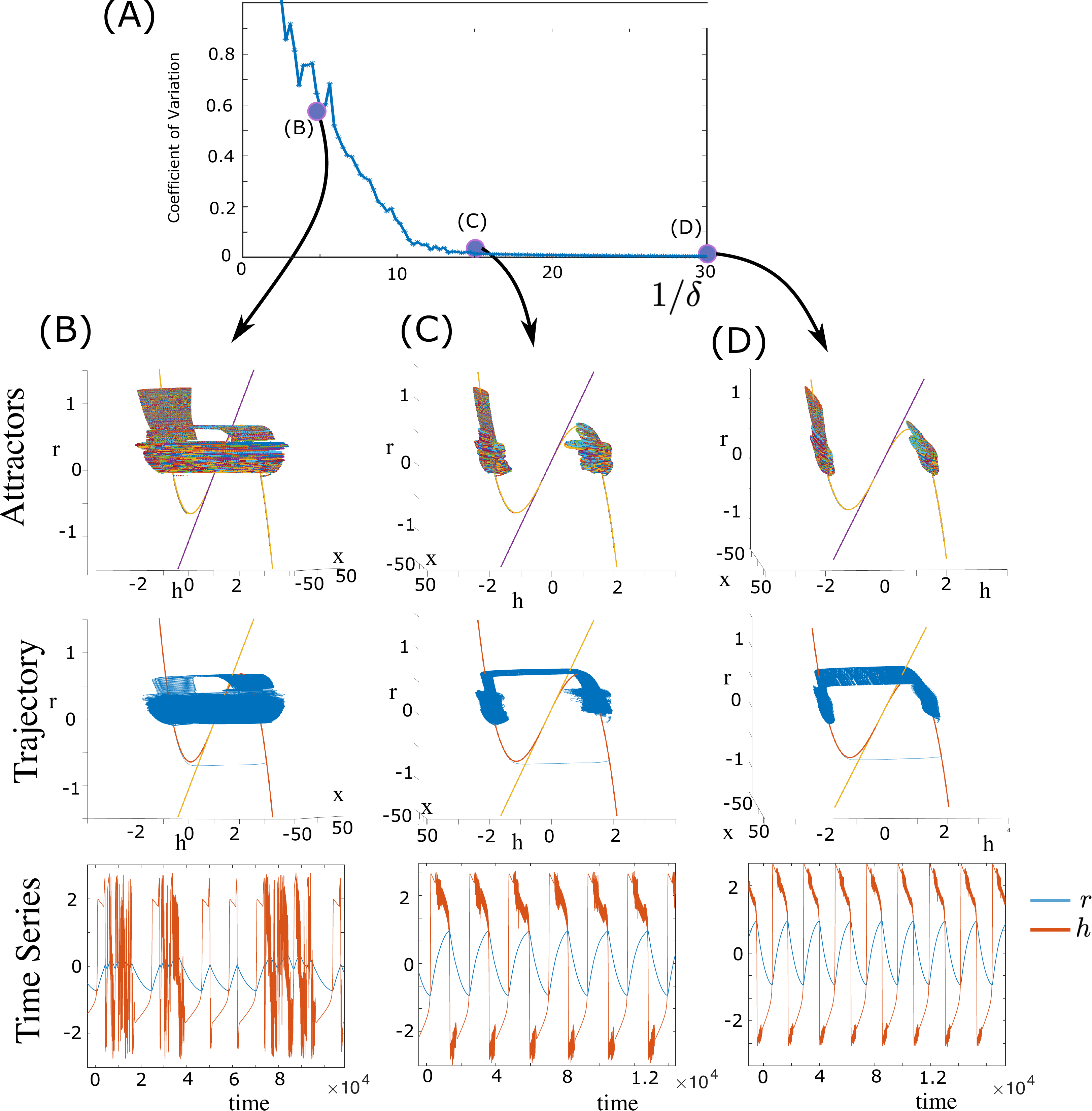}}
	\caption{Simulations of the FitzHugh-Nagumo Lorenz model, for $\delta=3$ (top,
a1,b1,c1) or $\delta=30$ (bottom, a2,b2,c2). Left: fast attractors,
where each color represents a distinct orbit (a1,a2), Middle: projection of the five-dimensional
trajectory of the system in the space $(r,h,x)$ (b1,b2) and Right:
time-evolution of the FitzHugh-Nagumo variables $r$ and $h$ (c1,c2). }
	\label{fig:lorenz}
\end{figure*}
The parameters
$\gamma$ and $\delta$ are scaling parameters that control the amplitude
of the coupling of those equations, with $\gamma$ fixed equal to $400$
allowing us to generate our type of dynamics, an $\delta$ being varied. The parameter
$\varepsilon=0.01$ is the slow timescale ratio, and the Lorenz system
parameters are fixed to $\sigma=10$ and $b=2.66$ throughout our
simulations. By design, the Lorenz system will generate fast chaotic
dynamics when the slow variable of the FitzHugh-Nagumo model $r$ is
large enough, which induces chaotic dynamics of the fast variable $h$.
The parameter $\delta$ controls the amplitude of the fast dynamics
chaotic attractor, and it can be chosen to induce an internal crisis.
For $\delta=1/30$ the chaotic attractors are significantly scaled-down
and remain far from the unstable fixed point (Fig.~\ref{fig:lorenz}D),
and the system generates relatively regular relaxation oscillations
through the chaotic attractors, corresponding to fast chaos. To estimate 
the fluctuations in the period of the relaxation cycles, we  have used a Poincaré section
placed on the hyperplane $h=0$ and recorded the crossings associated with $h$ 
increasing and $r<-0.2$. This condition was found to capture precisely the relaxation 
cycles away from the chaotic behaviors. For $\delta =1/30$, 
we confirmed that the fluctuations in the duration of the relaxation
cycles were very small compared to the mean, with a coefficient of
variation $c_v = 0.01$. 

As $\delta$ increases, we observe that the coefficient of variation
transitions from small values to significantly larger values, as the
fast attractor dynamics grows, generating a crisis and opening the
way to shortcut trajectories. These are visible in particular for
$\delta=1/4$ (Fig.~\ref{fig:lorenz}B), where the fast attractor clearly shows the presence of
two crises that open a shortcut avenue. Associated relaxation cycles
show an erratic alternation between trajectories that perform a full
relaxation cycle and those that transition earlier. 

\section{Discussion}

In this paper, we demonstrate that while chaos is usually associated with
disorder, the relationship between macroscopic patterns and chaos can be more
complex, and in particular mathematically chaotic systems can display
regular and robust dynamics at slow timescales. We  have shown that the presence of external crises of chaotic attractors combined with relaxation cycles predicted the existence of sharp
transitions between such relatively regular chaotic dynamics and erratic
behaviors at slow timescales. In these regimes, erratic alternations
between long and short bursts are governed by long irregular transients
around the ghost of a chaotic attractor, a scenario that we sharply
validated with the introduction of a purely probabilistic model from
which we predicted the frequency of occurrence of shorter bursts. Both
slow and fast chaos are  phenomena that arise in a variety of maps and
continuous systems. We showed that the coefficient of variation
associated with a time series of interburst intervals (or some other
appropriately chosen observable associated with the slow dynamics)
robustly distinguishes between slow and fast chaos in a variety of
scenarios, even including intracellular voltage data.
Our models yield dynamics very similar
to what is observed experimentally in the crab nervous system. We
further illustrate this with a 3-neuron model and various
quantifications of data and numerical simulations. 

This phenomenon adds to a wide
literature on chaotic dynamics of neural systems, that finely
characterized micro- and macro-chaotic structures in bifurcation
diagrams~\cite{barrio14} and the fine dynamics of chaotic alternations
in bursts at spike-adding transitions in slow-fast systems of
bursters~\cite{barrio21,medvedev,serrano21}, with a possible key role of
canards~\cite{shilnikov03}. Remarkable works also carefully investigated
the complex chaotic dynamics of models of pancreatic beta-cells
systems~\cite{duarte09}, and argued that another global bifurcation
called interior crises of chaotic attractors could be related to some
sudden changes in bursting behaviors~\cite{duarte17}. 

\begin{figure*}
\centerline{\includegraphics[width=.6\textwidth]{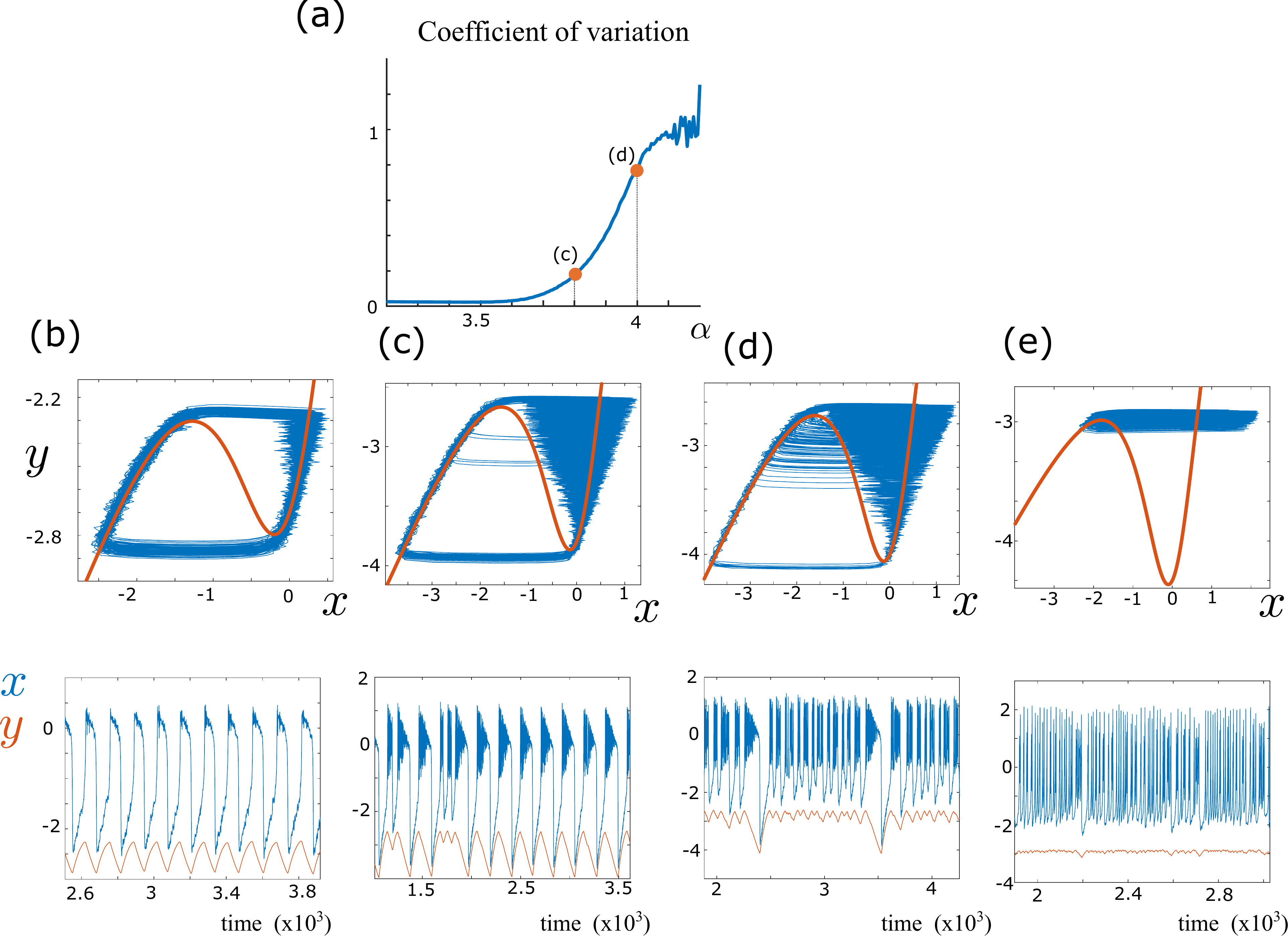}}
	\caption{Simulations of the Rulkov map with additive Gaussian noise on the voltage variable $x_{n+1} = \frac{\alpha}{1+x_n^2} +y_n +\gamma \xi_n$, $y_{n+1} = y_n - \mu \;  ( x_n - \sigma)$, with $\gamma=0.05$ and $(\xi_n)_{n\geq 0}$ independent standard Gaussian random variables. (a): coefficient of variation computed for $N=10^7$ iterates after a transient of $10^5$ iterates, with events computed as in Figure~\ref{fig:CoeffVar}. (b-e) trajectories in the phase plane (top) or as a function of time (bottom) for $\alpha=2.7$(b), $\alpha=3.8$ (c), $\alpha=4$ (d) and $\alpha=5$ (e). }
	\label{fig:Stochastic}
\end{figure*}

This paper is focused on deterministic chaos as a source of variability. 
Another interesting perspective of this work is the question of the
robustness of slow behaviors in stochastic systems and their breakdown. A
first observation is that the observations made in the Rulkov map
persisted in stochastic systems with small Gaussian additive noise: even
if theoretically shortcuts and changes of the period are allowed with
Gaussian noise, these occurred with rare probability, before the crisis,
and suddenly became much more frequent as the system approached the
deterministic crisis (see Figure~\ref{fig:Stochastic}). 

In stochastic systems 
however, chaos is not needed
to generate irregular behaviors and therefore such transitions between
fast and slow chaos may rely on a variety of other types of dynamical
and stochastic structures that would allow sudden switches in the
probability distribution of first passage times. The study of these
systems and structures open some new avenues of investigation.

\section*{Acknowledgements}
JJ was supported in part by NIH T32 NS007292. 
ES was partially supported by the Simons Foundation under Award 636383. JT acknowledges support from NSF DMS 1951369. 
JJ and ES acknowledge support from NSF grant DMS-1440140 while they were at 
residence at the Mathematical Sciences Research Institute in Berkeley, CA, during 
the Fall 2018 semester. S.K. was supported by funding from the National Institutes of Health (R35NS 097343).

\appendix
\section*{Appendix}
This Appendix gives a couple of details that are tangential to the main text. 
Section~\ref{sec:stg} gives a detailed description of the crustacean STG with data shown in Fig.~\ref{fig:data}.
 Section~\ref{sec:chaos} 
 qualitatively classifies the different types of chaos observed in the Rulkov map.  
 Section~\ref{sec:ComputationalMethodologies} discusses computational methodologies used 
 for the FitzHugh-Nagumo-Lorenz model.

\section{A detailed description of the STG model}\label{sec:stg}

The stomatogastric ganglion (STG) of crustaceans is a well-studied
rhythmic motor circuit that controls movements of muscles involved in
chewing and filtering of food. It produces a remarkably stable output in
the face of changing internal and external conditions and has provided
deep insights into the circuit and cellular-level mechanisms involved in
achieving circuit robustness and homeostasis. The pyloric circuit within
the STG is a central pattern generator that produces a characteristic
triphasic pattern of activity. A pair of pyloric dilator (PD) neurons
bursts initiate the rhythm, and they are followed in phase by a lateral
pyloric (LP) neuron and finally by the pyloric neurons (PYs). All
neurons have large amplitude slow-wave membrane voltage oscillations and
fire a burst of action potentials at the peak of the depolarization
during normal activity (Fig.~\ref{fig:data}a ). There is cycle-to-cycle variability of
minor features of this pattern, but the order of firing, relative phases, and duty cycles (fraction of the cycle period during which a neuron is
firing) are key to maintaining a functional output and are invariant.
Upon blocking of the neurohormonal inputs to the STG (decentralization),
this rhythmic bursting activity can change to an erratic spiking
behavior with sporadic bursts and a complete loss of rhythmicity. The
circuit is no longer functional in this state (Fig.~~\ref{fig:data}b). From the
mathematical viewpoint, STG circuit activity shows hallmarks of chaos
both in the rhythmic and non-rhythmic conditions, as shown in the
transition probability analysis plots in the bottom row of Fig.~\ref{fig:data}. 
Cycle-to-cycle variability of the spikes within a given burst,
especially in LP and PY spikes can be seen, which fills a region of space
in the three-variable plot even during normal triphasic activity. 

\section{Classification of chaotic behaviors}\label{sec:chaos}

The main text concentrates on distinguishing between fast and slow chaos, two qualitative behaviors
that appear most meaningful for qualitative descriptions. Our analyses
led us to further identify the following types of chaotic behaviors: 
 
{\em Weak chaos. } For a region of  $\alpha$ values 
 starting around $2.7$ we observe a phenomenon we refer to as  {\em weak chaos} cf.
Fig.~\ref{fig:Lyap_Traces}{a} in which the system is detected as chaotic using several measures of chaos, but does not noticeably alter the overall dynamics or functionality. The underlying mechanism 
 for this chaos appears to be referred to as a weakly chaotic ring in~\cite{mira05}, 
 the same as the phenomenon found in~\cite{lorenz1989}. 
 
 After $\alpha$ reaches the value $3$, the fast
 subsystem exhibits bistability, with a stable fixed point less than
 the resting membrane potential $\sigma$, and a chaotic attractor mostly 
 located above $\sigma$. Again, the time trajectories are not significantly altered
 in terms of functionality. 
 
   {\em Fast chaos.} 
 Starting at around $\alpha>3.5$, the system shows
 relaxation oscillations through the strange attractor of the fast
 variable, cf. Figs.~\ref{fig:Lyap_Traces}{b}.
 However, the decay
 of the slow variable takes a relatively fixed time. Thus 
 the interburst intervals 
 appear periodic at the
 slow timescale and the chaos is restricted to fast timescales. 
 
 {\em Slow chaos.}
 As $\alpha$ is increased towards 4, as in Fig.~\ref{fig:Lyap_Traces}{c}, 
 the full system escapes
 the chaotic region through a new route, or {\em shortcut}, distinct from the
 classical relaxation cycle. The plot of the slow variable shows that 
 it is unpredictable whether the
 escape from chaos will be through the classical or shortcut route. 
 Slow chaos  persists for $\alpha>4$, as in Fig.~\ref{fig:Lyap_Traces}{d}. 
 However, the frequency of the shortcut becomes greater as $\alpha$ increases.    
 Since slow chaos results in unpredictable interburst intervals, slow chaos results 
 in loss of functionality. 
 Terman~\cite{terman92} described slow chaos and symbolic dynamics in a neural model
 with relaxation cycles
 as a system transitions from periodic bursting to spiking, but the mechanism is 
 different and does not include the possibility of fast chaos.
 
{\em Hyperchaos.} When $\alpha$ is
 sufficiently away from the crisis, the frequency at which the system
 describes a full relaxation cycle vanishes, and the system fires
 repetitively without returning to the resting potential, 
 cf. Figs.~\ref{fig:Lyap_Traces}{e}.  The
 dynamics at $ \alpha$ above around $4.5$ displays {\em hyperchaos}, since it 
 has two positive Lyapunov exponents, cf. Fig.~\ref{fig:Lyap_Traces}{ bottom}; 
 see~\cite{rossler79,stankevich20}.

	\section{Computational Methodologies}
	\label{sec:ComputationalMethodologies}
	
	All of our numerical simulations were performed using code written in Matlab. 
	The source code for all numerical simulations is available at \cite{bib:codes}.
	
	Simulations of the continuous time FitzHugh-Nagumo-Lorenz model were performed in Matlab using the {\tt ode45} integrator with relative tolerance
	$10^{-5}$ and absolute tolerance $10^{-7}$. The trajectories for the 3 examples provided were
	confirmed using the routines {\tt ode113}, {\tt ode15s} and {\tt ode23s}, or with lower tolerances, and 
	we observed no qualitative difference in the trajectories. Solutions were computed over 
	$250\,000$ time units. Fast attractors were computed similarly over a range  
	of values of $r$ considered fixed, and starting with a range of initial conditions
	in the vicinity of the fixed points of the fast system (fixed points were computed
	using the Matlab {\tt vpasolve} function and initial conditions were drawn randomly
	according to a Gaussian centered the fixed point with standard deviation 0.01.


\begin{thebibliography}{49}%
\makeatletter
\providecommand \@ifxundefined [1]{%
 \@ifx{#1\undefined}
}%
\providecommand \@ifnum [1]{%
 \ifnum #1\expandafter \@firstoftwo
 \else \expandafter \@secondoftwo
 \fi
}%
\providecommand \@ifx [1]{%
 \ifx #1\expandafter \@firstoftwo
 \else \expandafter \@secondoftwo
 \fi
}%
\providecommand \natexlab [1]{#1}%
\providecommand \enquote  [1]{``#1''}%
\providecommand \bibnamefont  [1]{#1}%
\providecommand \bibfnamefont [1]{#1}%
\providecommand \citenamefont [1]{#1}%
\providecommand \href@noop [0]{\@secondoftwo}%
\providecommand \href [0]{\begingroup \@sanitize@url \@href}%
\providecommand \@href[1]{\@@startlink{#1}\@@href}%
\providecommand \@@href[1]{\endgroup#1\@@endlink}%
\providecommand \@sanitize@url [0]{\catcode `\\12\catcode `\$12\catcode
  `\&12\catcode `\#12\catcode `\^12\catcode `\_12\catcode `\%12\relax}%
\providecommand \@@startlink[1]{}%
\providecommand \@@endlink[0]{}%
\providecommand \url  [0]{\begingroup\@sanitize@url \@url }%
\providecommand \@url [1]{\endgroup\@href {#1}{\urlprefix }}%
\providecommand \urlprefix  [0]{URL }%
\providecommand \Eprint [0]{\href }%
\providecommand \doibase [0]{https://doi.org/}%
\providecommand \selectlanguage [0]{\@gobble}%
\providecommand \bibinfo  [0]{\@secondoftwo}%
\providecommand \bibfield  [0]{\@secondoftwo}%
\providecommand \translation [1]{[#1]}%
\providecommand \BibitemOpen [0]{}%
\providecommand \bibitemStop [0]{}%
\providecommand \bibitemNoStop [0]{.\EOS\space}%
\providecommand \EOS [0]{\spacefactor3000\relax}%
\providecommand \BibitemShut  [1]{\csname bibitem#1\endcsname}%
\let\auto@bib@innerbib\@empty
\bibitem [{\citenamefont {Hayashi}\ and\ \citenamefont
  {Ishizuka}(1992)}]{hayashi1992chaotic}%
  \BibitemOpen
  \bibfield  {author} {\bibinfo {author} {\bibfnamefont {H.}~\bibnamefont
  {Hayashi}}\ and\ \bibinfo {author} {\bibfnamefont {S.}~\bibnamefont
  {Ishizuka}},\ }\href@noop {} {\bibfield  {journal} {\bibinfo  {journal}
  {Journal of Theoretical Biology}\ }\textbf {\bibinfo {volume} {156}},\
  \bibinfo {pages} {269} (\bibinfo {year} {1992})}\BibitemShut {NoStop}%
\bibitem [{\citenamefont {Korn}\ and\ \citenamefont {Faure}(2003)}]{korn}%
  \BibitemOpen
  \bibfield  {author} {\bibinfo {author} {\bibfnamefont {H.}~\bibnamefont
  {Korn}}\ and\ \bibinfo {author} {\bibfnamefont {P.}~\bibnamefont {Faure}},\
  }\href@noop {} {\bibfield  {journal} {\bibinfo  {journal} {Comptes rendus
  biologies}\ }\textbf {\bibinfo {volume} {326}},\ \bibinfo {pages} {787}
  (\bibinfo {year} {2003})}\BibitemShut {NoStop}%
\bibitem [{\citenamefont {Kedia}\ and\ \citenamefont {Marder}(2022)}]{marder2}%
  \BibitemOpen
  \bibfield  {author} {\bibinfo {author} {\bibfnamefont {S.}~\bibnamefont
  {Kedia}}\ and\ \bibinfo {author} {\bibfnamefont {E.}~\bibnamefont {Marder}},\
  }\href@noop {} {\bibfield  {journal} {\bibinfo  {journal} {Current Biology}\
  }\textbf {\bibinfo {volume} {32}},\ \bibinfo {pages} {1439} (\bibinfo {year}
  {2022})}\BibitemShut {NoStop}%
\bibitem [{\citenamefont {Bernard}(1879)}]{bernard1879}%
  \BibitemOpen
  \bibfield  {author} {\bibinfo {author} {\bibfnamefont {C.}~\bibnamefont
  {Bernard}},\ }\href@noop {} {\emph {\bibinfo {title} {Le{\c{c}}ons sur les
  ph{\'e}nom{\`e}nes de la vie commune aux animaux et aux v{\'e}g{\'e}taux}}},\
  Vol.~\bibinfo {volume} {2}\ (\bibinfo  {publisher} {Bailli{\`e}re},\ \bibinfo
  {year} {1879})\BibitemShut {NoStop}%
\bibitem [{\citenamefont {May}\ \emph {et~al.}(1998)\citenamefont {May},
  \citenamefont {Vernoux}, \citenamefont {Leaver}, \citenamefont {Montagu},\
  and\ \citenamefont {Inz{\'e}}}]{may1998glutathione}%
  \BibitemOpen
  \bibfield  {author} {\bibinfo {author} {\bibfnamefont {M.~J.}\ \bibnamefont
  {May}}, \bibinfo {author} {\bibfnamefont {T.}~\bibnamefont {Vernoux}},
  \bibinfo {author} {\bibfnamefont {C.}~\bibnamefont {Leaver}}, \bibinfo
  {author} {\bibfnamefont {M.~V.}\ \bibnamefont {Montagu}},\ and\ \bibinfo
  {author} {\bibfnamefont {D.}~\bibnamefont {Inz{\'e}}},\ }\href@noop {}
  {\bibfield  {journal} {\bibinfo  {journal} {Journal of Experimental Botany}\
  }\textbf {\bibinfo {volume} {49}},\ \bibinfo {pages} {649} (\bibinfo {year}
  {1998})}\BibitemShut {NoStop}%
\bibitem [{\citenamefont {Dijk}\ \emph {et~al.}(1999)\citenamefont {Dijk},
  \citenamefont {Duffy}, \citenamefont {Riel}, \citenamefont {Shanahan},\ and\
  \citenamefont {Czeisler}}]{dijk1999ageing}%
  \BibitemOpen
  \bibfield  {author} {\bibinfo {author} {\bibfnamefont {D.-J.}\ \bibnamefont
  {Dijk}}, \bibinfo {author} {\bibfnamefont {J.~F.}\ \bibnamefont {Duffy}},
  \bibinfo {author} {\bibfnamefont {E.}~\bibnamefont {Riel}}, \bibinfo {author}
  {\bibfnamefont {T.~L.}\ \bibnamefont {Shanahan}},\ and\ \bibinfo {author}
  {\bibfnamefont {C.~A.}\ \bibnamefont {Czeisler}},\ }\href@noop {} {\bibfield
  {journal} {\bibinfo  {journal} {The Journal of physiology}\ }\textbf
  {\bibinfo {volume} {516}},\ \bibinfo {pages} {611} (\bibinfo {year}
  {1999})}\BibitemShut {NoStop}%
\bibitem [{\citenamefont {Aronoff}\ \emph {et~al.}(2004)\citenamefont
  {Aronoff}, \citenamefont {Berkowitz}, \citenamefont {Shreiner},\ and\
  \citenamefont {Want}}]{aronoff2004glucose}%
  \BibitemOpen
  \bibfield  {author} {\bibinfo {author} {\bibfnamefont {S.~L.}\ \bibnamefont
  {Aronoff}}, \bibinfo {author} {\bibfnamefont {K.}~\bibnamefont {Berkowitz}},
  \bibinfo {author} {\bibfnamefont {B.}~\bibnamefont {Shreiner}},\ and\
  \bibinfo {author} {\bibfnamefont {L.}~\bibnamefont {Want}},\ }\href@noop {}
  {\bibfield  {journal} {\bibinfo  {journal} {Diabetes spectrum}\ }\textbf
  {\bibinfo {volume} {17}},\ \bibinfo {pages} {183} (\bibinfo {year}
  {2004})}\BibitemShut {NoStop}%
\bibitem [{\citenamefont {Turrigiano}\ and\ \citenamefont
  {Nelson}(2004)}]{turrigiano2004homeostatic}%
  \BibitemOpen
  \bibfield  {author} {\bibinfo {author} {\bibfnamefont {G.~G.}\ \bibnamefont
  {Turrigiano}}\ and\ \bibinfo {author} {\bibfnamefont {S.~B.}\ \bibnamefont
  {Nelson}},\ }\href@noop {} {\bibfield  {journal} {\bibinfo  {journal} {Nature
  reviews neuroscience}\ }\textbf {\bibinfo {volume} {5}},\ \bibinfo {pages}
  {97} (\bibinfo {year} {2004})}\BibitemShut {NoStop}%
\bibitem [{\citenamefont {Cannon}\ and\ \citenamefont
  {Miller}(2016)}]{cannon2016synaptic}%
  \BibitemOpen
  \bibfield  {author} {\bibinfo {author} {\bibfnamefont {J.}~\bibnamefont
  {Cannon}}\ and\ \bibinfo {author} {\bibfnamefont {P.}~\bibnamefont
  {Miller}},\ }\href@noop {} {\bibfield  {journal} {\bibinfo  {journal}
  {Journal of neurophysiology}\ }\textbf {\bibinfo {volume} {116}},\ \bibinfo
  {pages} {2004} (\bibinfo {year} {2016})}\BibitemShut {NoStop}%
\bibitem [{\citenamefont {Marder}\ and\ \citenamefont
  {Tang}(2010)}]{marder2010coordinating}%
  \BibitemOpen
  \bibfield  {author} {\bibinfo {author} {\bibfnamefont {E.}~\bibnamefont
  {Marder}}\ and\ \bibinfo {author} {\bibfnamefont {L.~S.}\ \bibnamefont
  {Tang}},\ }\href@noop {} {\bibfield  {journal} {\bibinfo  {journal} {Neuron}\
  }\textbf {\bibinfo {volume} {66}},\ \bibinfo {pages} {161} (\bibinfo {year}
  {2010})}\BibitemShut {NoStop}%
\bibitem [{\citenamefont {Naud{\'e}}\ \emph {et~al.}(2013)\citenamefont
  {Naud{\'e}}, \citenamefont {Cessac}, \citenamefont {Berry},\ and\
  \citenamefont {Delord}}]{naude2013effects}%
  \BibitemOpen
  \bibfield  {author} {\bibinfo {author} {\bibfnamefont {J.}~\bibnamefont
  {Naud{\'e}}}, \bibinfo {author} {\bibfnamefont {B.}~\bibnamefont {Cessac}},
  \bibinfo {author} {\bibfnamefont {H.}~\bibnamefont {Berry}},\ and\ \bibinfo
  {author} {\bibfnamefont {B.}~\bibnamefont {Delord}},\ }\href@noop {}
  {\bibfield  {journal} {\bibinfo  {journal} {Journal of Neuroscience}\
  }\textbf {\bibinfo {volume} {33}},\ \bibinfo {pages} {15032} (\bibinfo {year}
  {2013})}\BibitemShut {NoStop}%
\bibitem [{\citenamefont {Goldberger}(1991)}]{goldberger1991normal}%
  \BibitemOpen
  \bibfield  {author} {\bibinfo {author} {\bibfnamefont {A.~L.}\ \bibnamefont
  {Goldberger}},\ }\href@noop {} {\bibfield  {journal} {\bibinfo  {journal}
  {Physiology}\ }\textbf {\bibinfo {volume} {6}},\ \bibinfo {pages} {87}
  (\bibinfo {year} {1991})}\BibitemShut {NoStop}%
\bibitem [{\citenamefont {Reed}\ \emph {et~al.}(2017)\citenamefont {Reed},
  \citenamefont {Best}, \citenamefont {Golubitsky}, \citenamefont {Stewart},\
  and\ \citenamefont {Nijhout}}]{reed2017analysis}%
  \BibitemOpen
  \bibfield  {author} {\bibinfo {author} {\bibfnamefont {M.}~\bibnamefont
  {Reed}}, \bibinfo {author} {\bibfnamefont {J.}~\bibnamefont {Best}}, \bibinfo
  {author} {\bibfnamefont {M.}~\bibnamefont {Golubitsky}}, \bibinfo {author}
  {\bibfnamefont {I.}~\bibnamefont {Stewart}},\ and\ \bibinfo {author}
  {\bibfnamefont {H.~F.}\ \bibnamefont {Nijhout}},\ }\href@noop {} {\bibfield
  {journal} {\bibinfo  {journal} {Bulletin of mathematical biology}\ }\textbf
  {\bibinfo {volume} {79}},\ \bibinfo {pages} {2534} (\bibinfo {year}
  {2017})}\BibitemShut {NoStop}%
\bibitem [{\citenamefont {Golubitsky}\ and\ \citenamefont
  {Stewart}(2018)}]{golubitsky2018homeostasis}%
  \BibitemOpen
  \bibfield  {author} {\bibinfo {author} {\bibfnamefont {M.}~\bibnamefont
  {Golubitsky}}\ and\ \bibinfo {author} {\bibfnamefont {I.}~\bibnamefont
  {Stewart}},\ }\href@noop {} {\bibfield  {journal} {\bibinfo  {journal} {SIAM
  Journal on Applied Dynamical Systems}\ }\textbf {\bibinfo {volume} {17}},\
  \bibinfo {pages} {1816} (\bibinfo {year} {2018})}\BibitemShut {NoStop}%
\bibitem [{\citenamefont {Golubitsky}\ and\ \citenamefont
  {Wang}(2020)}]{golubitsky2020infinitesimal}%
  \BibitemOpen
  \bibfield  {author} {\bibinfo {author} {\bibfnamefont {M.}~\bibnamefont
  {Golubitsky}}\ and\ \bibinfo {author} {\bibfnamefont {Y.}~\bibnamefont
  {Wang}},\ }\href@noop {} {\bibfield  {journal} {\bibinfo  {journal} {Journal
  of mathematical biology}\ }\textbf {\bibinfo {volume} {80}},\ \bibinfo
  {pages} {1163} (\bibinfo {year} {2020})}\BibitemShut {NoStop}%
\bibitem [{\citenamefont {Yu}\ and\ \citenamefont {Thomas}(2022)}]{thomas}%
  \BibitemOpen
  \bibfield  {author} {\bibinfo {author} {\bibfnamefont {Z.}~\bibnamefont
  {Yu}}\ and\ \bibinfo {author} {\bibfnamefont {P.~J.}\ \bibnamefont
  {Thomas}},\ }\href@noop {} {\bibfield  {journal} {\bibinfo  {journal}
  {Journal of Mathematical Biology}\ }\textbf {\bibinfo {volume} {84}},\
  \bibinfo {pages} {1} (\bibinfo {year} {2022})}\BibitemShut {NoStop}%
\bibitem [{\citenamefont {Marder}\ and\ \citenamefont
  {Bucher}(2007)}]{marder:07}%
  \BibitemOpen
  \bibfield  {author} {\bibinfo {author} {\bibfnamefont {E.}~\bibnamefont
  {Marder}}\ and\ \bibinfo {author} {\bibfnamefont {D.}~\bibnamefont
  {Bucher}},\ }\href@noop {} {\bibfield  {journal} {\bibinfo  {journal} {Annu.
  Rev. Physiol.}\ }\textbf {\bibinfo {volume} {69}},\ \bibinfo {pages} {291}
  (\bibinfo {year} {2007})}\BibitemShut {NoStop}%
\bibitem [{\citenamefont {Alonso}\ and\ \citenamefont {Marder}(2020)}]{alonso}%
  \BibitemOpen
  \bibfield  {author} {\bibinfo {author} {\bibfnamefont {L.~M.}\ \bibnamefont
  {Alonso}}\ and\ \bibinfo {author} {\bibfnamefont {E.}~\bibnamefont
  {Marder}},\ }\href@noop {} {\bibfield  {journal} {\bibinfo  {journal}
  {Elife}\ }\textbf {\bibinfo {volume} {9}},\ \bibinfo {pages} {e55470}
  (\bibinfo {year} {2020})}\BibitemShut {NoStop}%
\bibitem [{\citenamefont {Alonso}\ and\ \citenamefont
  {Marder}(2019)}]{alonso2}%
  \BibitemOpen
  \bibfield  {author} {\bibinfo {author} {\bibfnamefont {L.~M.}\ \bibnamefont
  {Alonso}}\ and\ \bibinfo {author} {\bibfnamefont {E.}~\bibnamefont
  {Marder}},\ }\href@noop {} {\bibfield  {journal} {\bibinfo  {journal}
  {Elife}\ }\textbf {\bibinfo {volume} {8}} (\bibinfo {year}
  {2019})}\BibitemShut {NoStop}%
\bibitem [{\citenamefont {Sander}\ and\ \citenamefont
  {Yorke}(2015)}]{sander2015many}%
  \BibitemOpen
  \bibfield  {author} {\bibinfo {author} {\bibfnamefont {E.}~\bibnamefont
  {Sander}}\ and\ \bibinfo {author} {\bibfnamefont {J.~A.}\ \bibnamefont
  {Yorke}},\ }\href@noop {} {\bibfield  {journal} {\bibinfo  {journal}
  {International Journal of Bifurcation and Chaos}\ }\textbf {\bibinfo {volume}
  {25}},\ \bibinfo {pages} {1530011} (\bibinfo {year} {2015})}\BibitemShut
  {NoStop}%
\bibitem [{\citenamefont {Jaquette}\ \emph {et~al.}(2023)\citenamefont
  {Jaquette}, \citenamefont {Kedia}, \citenamefont {Sander},\ and\
  \citenamefont {Touboul}}]{bib:codes}%
  \BibitemOpen
  \bibfield  {author} {\bibinfo {author} {\bibfnamefont {J.}~\bibnamefont
  {Jaquette}}, \bibinfo {author} {\bibfnamefont {S.}~\bibnamefont {Kedia}},
  \bibinfo {author} {\bibfnamefont {E.}~\bibnamefont {Sander}},\ and\ \bibinfo
  {author} {\bibfnamefont {J.~D.}\ \bibnamefont {Touboul}},\ }\href@noop {}
  {\bibinfo {title} {Codes of ``{R}eliability and robustness of oscillations in
  some slow-fast chaotic systems''}},\ \bibinfo {howpublished}
  {\url{https://github.com/JCJaquette/Slow-Fast-Chaos}} (\bibinfo {year}
  {2023})\BibitemShut {NoStop}%
\bibitem [{\citenamefont {Maslennikov}\ and\ \citenamefont
  {Nekorkin}(2016)}]{maslennikov16}%
  \BibitemOpen
  \bibfield  {author} {\bibinfo {author} {\bibfnamefont {O.~V.}\ \bibnamefont
  {Maslennikov}}\ and\ \bibinfo {author} {\bibfnamefont {V.~I.}\ \bibnamefont
  {Nekorkin}},\ }\href {https://doi.org/10.1063/1.4955084} {\bibfield
  {journal} {\bibinfo  {journal} {Chaos}\ }\textbf {\bibinfo {volume} {26}},\
  \bibinfo {pages} {073104, 10} (\bibinfo {year} {2016})}\BibitemShut {NoStop}%
\bibitem [{\citenamefont {Han}\ \emph {et~al.}(2017)\citenamefont {Han},
  \citenamefont {Zhang}, \citenamefont {Yu},\ and\ \citenamefont {Bi}}]{han17}%
  \BibitemOpen
  \bibfield  {author} {\bibinfo {author} {\bibfnamefont {X.}~\bibnamefont
  {Han}}, \bibinfo {author} {\bibfnamefont {C.}~\bibnamefont {Zhang}}, \bibinfo
  {author} {\bibfnamefont {Y.}~\bibnamefont {Yu}},\ and\ \bibinfo {author}
  {\bibfnamefont {Q.}~\bibnamefont {Bi}},\ }\href
  {https://doi.org/10.1142/S0218127417500511} {\bibfield  {journal} {\bibinfo
  {journal} {Internat. J. Bifur. Chaos Appl. Sci. Engrg.}\ }\textbf {\bibinfo
  {volume} {27}},\ \bibinfo {pages} {1750051, 17} (\bibinfo {year}
  {2017})}\BibitemShut {NoStop}%
\bibitem [{\citenamefont {Terman}(1992)}]{terman92}%
  \BibitemOpen
  \bibfield  {author} {\bibinfo {author} {\bibfnamefont {D.}~\bibnamefont
  {Terman}},\ }\href {https://doi.org/10.1007/BF02429854} {\bibfield  {journal}
  {\bibinfo  {journal} {J. Nonlinear Sci.}\ }\textbf {\bibinfo {volume} {2}},\
  \bibinfo {pages} {135} (\bibinfo {year} {1992})}\BibitemShut {NoStop}%
\bibitem [{\citenamefont {Rulkov}(2001)}]{rulkov2001regularization}%
  \BibitemOpen
  \bibfield  {author} {\bibinfo {author} {\bibfnamefont {N.~F.}\ \bibnamefont
  {Rulkov}},\ }\href@noop {} {\bibfield  {journal} {\bibinfo  {journal}
  {Physical Review Letters}\ }\textbf {\bibinfo {volume} {86}},\ \bibinfo
  {pages} {183} (\bibinfo {year} {2001})}\BibitemShut {NoStop}%
\bibitem [{\citenamefont {Ibarz}\ \emph {et~al.}(2011)\citenamefont {Ibarz},
  \citenamefont {Casado},\ and\ \citenamefont {Sanju{\'a}n}}]{ibarz2011map}%
  \BibitemOpen
  \bibfield  {author} {\bibinfo {author} {\bibfnamefont {B.}~\bibnamefont
  {Ibarz}}, \bibinfo {author} {\bibfnamefont {J.~M.}\ \bibnamefont {Casado}},\
  and\ \bibinfo {author} {\bibfnamefont {M.~A.}\ \bibnamefont {Sanju{\'a}n}},\
  }\href@noop {} {\bibfield  {journal} {\bibinfo  {journal} {Physics reports}\
  }\textbf {\bibinfo {volume} {501}},\ \bibinfo {pages} {1} (\bibinfo {year}
  {2011})}\BibitemShut {NoStop}%
\bibitem [{\citenamefont {Courbage}\ and\ \citenamefont
  {Nekorkin}(2010)}]{courbage2010map}%
  \BibitemOpen
  \bibfield  {author} {\bibinfo {author} {\bibfnamefont {M.}~\bibnamefont
  {Courbage}}\ and\ \bibinfo {author} {\bibfnamefont {V.~I.}\ \bibnamefont
  {Nekorkin}},\ }\href@noop {} {\bibfield  {journal} {\bibinfo  {journal}
  {International Journal of Bifurcation and Chaos}\ }\textbf {\bibinfo {volume}
  {20}},\ \bibinfo {pages} {1631} (\bibinfo {year} {2010})}\BibitemShut
  {NoStop}%
\bibitem [{Note1()}]{Note1}%
  \BibitemOpen
  \bibinfo {note} {This cycle emerges through a Neimark-Sacker bifurcation
  arising at $\alpha = -(1-\mu )(1+\sigma ^2)^2/(2 \sigma )$ for almost every
  $\mu <3$ (for our parameters, $\alpha $ slightly below 2).}\BibitemShut
  {Stop}%
\bibitem [{\citenamefont {Grebogi}\ \emph {et~al.}(1983)\citenamefont
  {Grebogi}, \citenamefont {Ott},\ and\ \citenamefont {Yorke}}]{grebogi:83}%
  \BibitemOpen
  \bibfield  {author} {\bibinfo {author} {\bibfnamefont {C.}~\bibnamefont
  {Grebogi}}, \bibinfo {author} {\bibfnamefont {E.}~\bibnamefont {Ott}},\ and\
  \bibinfo {author} {\bibfnamefont {J.~A.}\ \bibnamefont {Yorke}},\ }\href@noop
  {} {\bibfield  {journal} {\bibinfo  {journal} {Physica D: Nonlinear
  Phenomena}\ }\textbf {\bibinfo {volume} {7}} (\bibinfo {year}
  {1983})}\BibitemShut {NoStop}%
\bibitem [{\citenamefont {Grebogi}\ \emph {et~al.}(1985)\citenamefont
  {Grebogi}, \citenamefont {Ott},\ and\ \citenamefont {Yorke}}]{grebogi:85}%
  \BibitemOpen
  \bibfield  {author} {\bibinfo {author} {\bibfnamefont {C.}~\bibnamefont
  {Grebogi}}, \bibinfo {author} {\bibfnamefont {E.}~\bibnamefont {Ott}},\ and\
  \bibinfo {author} {\bibfnamefont {J.~A.}\ \bibnamefont {Yorke}},\ }\href@noop
  {} {\bibfield  {journal} {\bibinfo  {journal} {Ergodic Theory and Dynamical
  Systems}\ }\textbf {\bibinfo {volume} {5}},\ \bibinfo {pages} {341} (\bibinfo
  {year} {1985})}\BibitemShut {NoStop}%
\bibitem [{\citenamefont {Grebogi}\ \emph {et~al.}(1987)\citenamefont
  {Grebogi}, \citenamefont {Ott}, \citenamefont {Romeiras},\ and\ \citenamefont
  {Yorke}}]{grebogi:87}%
  \BibitemOpen
  \bibfield  {author} {\bibinfo {author} {\bibfnamefont {C.}~\bibnamefont
  {Grebogi}}, \bibinfo {author} {\bibfnamefont {E.}~\bibnamefont {Ott}},
  \bibinfo {author} {\bibfnamefont {F.}~\bibnamefont {Romeiras}},\ and\
  \bibinfo {author} {\bibfnamefont {J.~A.}\ \bibnamefont {Yorke}},\ }\href@noop
  {} {\bibfield  {journal} {\bibinfo  {journal} {Physical Review A}\ }\textbf
  {\bibinfo {volume} {36}},\ \bibinfo {pages} {5365} (\bibinfo {year}
  {1987})}\BibitemShut {NoStop}%
\bibitem [{\citenamefont {Gedeon}\ \emph {et~al.}(1999)\citenamefont {Gedeon},
  \citenamefont {Kokubu}, \citenamefont {Mischaikow}, \citenamefont {Oka},\
  and\ \citenamefont {Reineck}}]{gedeon}%
  \BibitemOpen
  \bibfield  {author} {\bibinfo {author} {\bibfnamefont {T.}~\bibnamefont
  {Gedeon}}, \bibinfo {author} {\bibfnamefont {H.}~\bibnamefont {Kokubu}},
  \bibinfo {author} {\bibfnamefont {K.}~\bibnamefont {Mischaikow}}, \bibinfo
  {author} {\bibfnamefont {H.}~\bibnamefont {Oka}},\ and\ \bibinfo {author}
  {\bibfnamefont {J.~F.}\ \bibnamefont {Reineck}},\ }\href@noop {} {\bibfield
  {journal} {\bibinfo  {journal} {Journal of Dynamics and Differential
  Equations}\ }\textbf {\bibinfo {volume} {11}},\ \bibinfo {pages} {427}
  (\bibinfo {year} {1999})}\BibitemShut {NoStop}%
\bibitem [{\citenamefont {Yorke}\ and\ \citenamefont
  {Alligood}(1985)}]{alligood-yorke}%
  \BibitemOpen
  \bibfield  {author} {\bibinfo {author} {\bibfnamefont {J.~A.}\ \bibnamefont
  {Yorke}}\ and\ \bibinfo {author} {\bibfnamefont {K.~T.}\ \bibnamefont
  {Alligood}},\ }\href@noop {} {\bibfield  {journal} {\bibinfo  {journal}
  {Communications in mathematical physics}\ }\textbf {\bibinfo {volume}
  {101}},\ \bibinfo {pages} {305} (\bibinfo {year} {1985})}\BibitemShut
  {NoStop}%
\bibitem [{Note2()}]{Note2}%
  \BibitemOpen
  \bibinfo {note} {Here, we chose \protect \[-\protect \frac 1 2 (\protect
  \qopname \relax o{tanh}(x-\theta )+\protect \qopname \relax o{tanh}(-x-\theta
  ))\protect \] with $\theta =2$.}\BibitemShut {Stop}%
\bibitem [{\citenamefont {FitzHugh}(1955)}]{fitzhugh1}%
  \BibitemOpen
  \bibfield  {author} {\bibinfo {author} {\bibfnamefont {R.}~\bibnamefont
  {FitzHugh}},\ }\href@noop {} {\bibfield  {journal} {\bibinfo  {journal}
  {Bulletin of Mathematical Biophysics}\ }\textbf {\bibinfo {volume} {17}},\
  \bibinfo {pages} {257} (\bibinfo {year} {1955})}\BibitemShut {NoStop}%
\bibitem [{\citenamefont {FitzHugh}(1961)}]{fitzhugh2}%
  \BibitemOpen
  \bibfield  {author} {\bibinfo {author} {\bibfnamefont {R.}~\bibnamefont
  {FitzHugh}},\ }\href@noop {} {\bibfield  {journal} {\bibinfo  {journal}
  {Biophysical Journal}\ }\textbf {\bibinfo {volume} {1}},\ \bibinfo {pages}
  {445} (\bibinfo {year} {1961})}\BibitemShut {NoStop}%
\bibitem [{\citenamefont {Nagumo}\ \emph {et~al.}(1962)\citenamefont {Nagumo},
  \citenamefont {Arimoto},\ and\ \citenamefont {Yoshizawa}}]{nagumo}%
  \BibitemOpen
  \bibfield  {author} {\bibinfo {author} {\bibfnamefont {J.}~\bibnamefont
  {Nagumo}}, \bibinfo {author} {\bibfnamefont {S.}~\bibnamefont {Arimoto}},\
  and\ \bibinfo {author} {\bibfnamefont {S.}~\bibnamefont {Yoshizawa}},\
  }\href@noop {} {\bibfield  {journal} {\bibinfo  {journal} {Proceedings of the
  IRE}\ }\textbf {\bibinfo {volume} {50}},\ \bibinfo {pages} {2061} (\bibinfo
  {year} {1962})}\BibitemShut {NoStop}%
\bibitem [{\citenamefont {Lorenz}(1963)}]{lorenz}%
  \BibitemOpen
  \bibfield  {author} {\bibinfo {author} {\bibfnamefont {E.~N.}\ \bibnamefont
  {Lorenz}},\ }\href@noop {} {\bibfield  {journal} {\bibinfo  {journal}
  {Journal of the Atmospheric Sciences}\ }\textbf {\bibinfo {volume} {20}},\
  \bibinfo {pages} {130} (\bibinfo {year} {1963})}\BibitemShut {NoStop}%
\bibitem [{\citenamefont {Barrio}\ \emph {et~al.}(2014)\citenamefont {Barrio},
  \citenamefont {Mart\'{\i}nez}, \citenamefont {Serrano},\ and\ \citenamefont
  {Shilnikov}}]{barrio14}%
  \BibitemOpen
  \bibfield  {author} {\bibinfo {author} {\bibfnamefont {R.}~\bibnamefont
  {Barrio}}, \bibinfo {author} {\bibfnamefont {M.~A.}\ \bibnamefont
  {Mart\'{\i}nez}}, \bibinfo {author} {\bibfnamefont {S.}~\bibnamefont
  {Serrano}},\ and\ \bibinfo {author} {\bibfnamefont {A.}~\bibnamefont
  {Shilnikov}},\ }\href {https://doi.org/10.1063/1.4882171} {\bibfield
  {journal} {\bibinfo  {journal} {Chaos}\ }\textbf {\bibinfo {volume} {24}},\
  \bibinfo {pages} {023128, 11} (\bibinfo {year} {2014})}\BibitemShut {NoStop}%
\bibitem [{\citenamefont {Barrio}\ \emph {et~al.}(2021)\citenamefont {Barrio},
  \citenamefont {Ib\'{a}\~{n}ez}, \citenamefont {P\'{e}rez},\ and\
  \citenamefont {Serrano}}]{barrio21}%
  \BibitemOpen
  \bibfield  {author} {\bibinfo {author} {\bibfnamefont {R.}~\bibnamefont
  {Barrio}}, \bibinfo {author} {\bibfnamefont {S.}~\bibnamefont
  {Ib\'{a}\~{n}ez}}, \bibinfo {author} {\bibfnamefont {L.}~\bibnamefont
  {P\'{e}rez}},\ and\ \bibinfo {author} {\bibfnamefont {S.}~\bibnamefont
  {Serrano}},\ }\href {https://doi.org/10.1063/5.0037942} {\bibfield  {journal}
  {\bibinfo  {journal} {Chaos}\ }\textbf {\bibinfo {volume} {31}},\ \bibinfo
  {pages} {043120, 14} (\bibinfo {year} {2021})}\BibitemShut {NoStop}%
\bibitem [{\citenamefont {Medvedev}(2006)}]{medvedev}%
  \BibitemOpen
  \bibfield  {author} {\bibinfo {author} {\bibfnamefont {G.~S.}\ \bibnamefont
  {Medvedev}},\ }\href@noop {} {\bibfield  {journal} {\bibinfo  {journal}
  {PRL}\ }\textbf {\bibinfo {volume} {97}} (\bibinfo {year}
  {2006})}\BibitemShut {NoStop}%
\bibitem [{\citenamefont {Serrano}\ \emph {et~al.}(2021)\citenamefont
  {Serrano}, \citenamefont {Mart\'{\i}nez},\ and\ \citenamefont
  {Barrio}}]{serrano21}%
  \BibitemOpen
  \bibfield  {author} {\bibinfo {author} {\bibfnamefont {S.}~\bibnamefont
  {Serrano}}, \bibinfo {author} {\bibfnamefont {M.~A.}\ \bibnamefont
  {Mart\'{\i}nez}},\ and\ \bibinfo {author} {\bibfnamefont {R.}~\bibnamefont
  {Barrio}},\ }\href {https://doi.org/10.1063/5.0043302} {\bibfield  {journal}
  {\bibinfo  {journal} {Chaos}\ }\textbf {\bibinfo {volume} {31}},\ \bibinfo
  {pages} {043108, 25} (\bibinfo {year} {2021})}\BibitemShut {NoStop}%
\bibitem [{\citenamefont {Shilnikov}\ and\ \citenamefont
  {Rulkov}(2003)}]{shilnikov03}%
  \BibitemOpen
  \bibfield  {author} {\bibinfo {author} {\bibfnamefont {A.~L.}\ \bibnamefont
  {Shilnikov}}\ and\ \bibinfo {author} {\bibfnamefont {N.~F.}\ \bibnamefont
  {Rulkov}},\ }\href {https://doi.org/10.1142/S0218127403008521} {\bibfield
  {journal} {\bibinfo  {journal} {Internat. J. Bifur. Chaos Appl. Sci. Engrg.}\
  }\textbf {\bibinfo {volume} {13}},\ \bibinfo {pages} {3325} (\bibinfo {year}
  {2003})}\BibitemShut {NoStop}%
\bibitem [{\citenamefont {Duarte}\ \emph {et~al.}(2009)\citenamefont {Duarte},
  \citenamefont {Janu\'{a}rio},\ and\ \citenamefont {Martins}}]{duarte09}%
  \BibitemOpen
  \bibfield  {author} {\bibinfo {author} {\bibfnamefont {J.}~\bibnamefont
  {Duarte}}, \bibinfo {author} {\bibfnamefont {C.}~\bibnamefont
  {Janu\'{a}rio}},\ and\ \bibinfo {author} {\bibfnamefont {N.}~\bibnamefont
  {Martins}},\ }\href {https://doi.org/10.1016/j.physd.2009.08.010} {\bibfield
  {journal} {\bibinfo  {journal} {Phys. D}\ }\textbf {\bibinfo {volume}
  {238}},\ \bibinfo {pages} {2129} (\bibinfo {year} {2009})}\BibitemShut
  {NoStop}%
\bibitem [{\citenamefont {Duarte}\ \emph {et~al.}(2017)\citenamefont {Duarte},
  \citenamefont {Janu\'{a}rio},\ and\ \citenamefont {Martins}}]{duarte17}%
  \BibitemOpen
  \bibfield  {author} {\bibinfo {author} {\bibfnamefont {J.}~\bibnamefont
  {Duarte}}, \bibinfo {author} {\bibfnamefont {C.}~\bibnamefont
  {Janu\'{a}rio}},\ and\ \bibinfo {author} {\bibfnamefont {N.}~\bibnamefont
  {Martins}},\ }\href {https://doi.org/10.3934/mbe.2017045} {\bibfield
  {journal} {\bibinfo  {journal} {Math. Biosci. Eng.}\ }\textbf {\bibinfo
  {volume} {14}},\ \bibinfo {pages} {821} (\bibinfo {year} {2017})}\BibitemShut
  {NoStop}%
\bibitem [{\citenamefont {Mira}\ and\ \citenamefont
  {Shilnikov}(2005)}]{mira05}%
  \BibitemOpen
  \bibfield  {author} {\bibinfo {author} {\bibfnamefont {C.}~\bibnamefont
  {Mira}}\ and\ \bibinfo {author} {\bibfnamefont {A.}~\bibnamefont
  {Shilnikov}},\ }\href {https://doi.org/10.1142/S0218127405014192} {\bibfield
  {journal} {\bibinfo  {journal} {Internat. J. Bifur. Chaos Appl. Sci. Engrg.}\
  }\textbf {\bibinfo {volume} {15}},\ \bibinfo {pages} {3509} (\bibinfo {year}
  {2005})}\BibitemShut {NoStop}%
\bibitem [{\citenamefont {Lorenz}(1989)}]{lorenz1989}%
  \BibitemOpen
  \bibfield  {author} {\bibinfo {author} {\bibfnamefont {E.~N.}\ \bibnamefont
  {Lorenz}},\ }\href@noop {} {\bibfield  {journal} {\bibinfo  {journal} {Phys.
  D}\ }\textbf {\bibinfo {volume} {35}},\ \bibinfo {pages} {299} (\bibinfo
  {year} {1989})}\BibitemShut {NoStop}%
\bibitem [{\citenamefont {R\"{o}ssler}(1979)}]{rossler79}%
  \BibitemOpen
  \bibfield  {author} {\bibinfo {author} {\bibfnamefont {O.~E.}\ \bibnamefont
  {R\"{o}ssler}},\ }\href {https://doi.org/10.1016/0375-9601(79)90150-6}
  {\bibfield  {journal} {\bibinfo  {journal} {Phys. Lett. A}\ }\textbf
  {\bibinfo {volume} {71}},\ \bibinfo {pages} {155} (\bibinfo {year}
  {1979})}\BibitemShut {NoStop}%
\bibitem [{\citenamefont {Stankevich}\ \emph {et~al.}(2021)\citenamefont
  {Stankevich}, \citenamefont {Kazakov},\ and\ \citenamefont
  {Gonchenko}}]{stankevich20}%
  \BibitemOpen
  \bibfield  {author} {\bibinfo {author} {\bibfnamefont {N.}~\bibnamefont
  {Stankevich}}, \bibinfo {author} {\bibfnamefont {A.}~\bibnamefont
  {Kazakov}},\ and\ \bibinfo {author} {\bibfnamefont {S.}~\bibnamefont
  {Gonchenko}},\ }\href {https://doi.org/10.1063/5.0050186} {\bibfield
  {journal} {\bibinfo  {journal} {Chaos}\ }\textbf {\bibinfo {volume} {31}},\
  \bibinfo {pages} {049903, 1} (\bibinfo {year} {2021})}\BibitemShut {NoStop}%
\end{thebibliography}

%

\end{document}